\documentclass[manuscript,screen]{acmart}
\renewcommand\footnotetextcopyrightpermission[1]{} 
\AtBeginDocument{%
  }
\usepackage{graphicx}
\usepackage{amsmath}
\usepackage{tabularx}
\usepackage{amsfonts}
\usepackage{url}
\usepackage{svg}
\usepackage{verbatim}
\usepackage{booktabs}
\usepackage{pifont}
\usepackage{times,epsfig}
\usepackage{makecell}
\usepackage{graphicx}
\usepackage{diagbox}
\usepackage[normalem]{ulem}
\usepackage{multirow}
\usepackage{graphicx}
\usepackage[normalem]{ulem}
\useunder{\uline}{\ul}{}
\usepackage{enumitem}
\usepackage{psfrag}

\usepackage{subfigure}
\usepackage{stfloats}
\usepackage{setspace}
\usepackage{wasysym}
\usepackage{multirow}
\usepackage{colortbl}
\usepackage[justification=centering]{caption}
\usepackage{fancyhdr}
\usepackage{comment}
\usepackage{xcolor}
\usepackage{tikz}
\usetikzlibrary{shapes.geometric, arrows, positioning, fit, calc}

\usepackage{xcolor}
\begin{document}

\title{Forewarned is Forearmed: A Survey on Large Language Model-based Agents in Autonomous Cyberattacks}

\author{Minrui Xu}
\email{MINRUI001@e.ntu.edu.sg}
\affiliation{%
  \institution{Nanyang Technological University}
  \country{Singapore}
}
\author{Jiani Fan}
\email{jiani001@e.ntu.edu.sg}
\affiliation{%
  \institution{Nanyang Technological University}
  \country{Singapore}
}
\author{Xinyu Huang}
\email{x357huan@uwaterloo.ca}
\affiliation{%
  \institution{University of Waterloo}
  \city{Waterloo}
  \state{ON}
  \country{Canada}
}
\author{Conghao Zhou}
\email{c89zhou@uwaterloo.ca}
\affiliation{%
  \institution{University of Waterloo}
  \city{Waterloo}
  \state{ON}
  \country{Canada}
}
\author{Jiawen Kang}
\email{kavinkang@gdut.edu.cn}
\affiliation{%
  \institution{Guangdong University of Technology}
  \city{Guangzhou}
  \state{Guangdong}
  \country{China}
}
\author{Dusit Niyato}
\email{dniyato@ntu.edu.sg}
\affiliation{%
  \institution{Nanyang Technological University}
  \country{Singapore}
}
\author{Shiwen Mao}
\email{smao@ieee.org}
\affiliation{%
  \institution{Auburn University}
  \country{USA}
}
\author{Zhu Han}
\email{hanzhu22@gmail.com}
\affiliation{%
  \institution{University of Houston}
  \city{Houston}
  \state{TX}
  \country{USA}
}
\author{Xuemin (Sherman) Shen}
\email{sshen@uwaterloo.ca}
\affiliation{%
  \institution{University of Waterloo}
  \city{Waterloo}
  \state{ON}
  \country{Canada}
}
\author{Kwok-Yan Lam}
\email{kwokyan.lam@ntu.edu.sg}
\affiliation{%
  \institution{Nanyang Technological University}
  \country{Singapore}
}
\renewcommand{\shortauthors}{Xu et al.}

\begin{abstract}
With the continuous evolution of Large Language Models (LLMs), LLM-based agents have advanced beyond passive chatbots to become autonomous cyber entities capable of performing complex tasks, including web browsing, malicious code and deceptive content generation, and decision-making. By significantly reducing the time, expertise, and resources, AI-assisted cyberattacks orchestrated by LLM-based agents have led to a phenomenon termed Cyber Threat Inflation, characterized by a significant reduction in attack costs and a tremendous increase in attack scale. To provide actionable defensive insights, in this survey, we focus on the potential cyber threats posed by LLM-based agents across diverse network systems. 
Firstly, we present the capabilities of LLM-based cyberattack agents, which include executing autonomous attack strategies, comprising scouting, memory, reasoning, and action, and facilitating collaborative operations with other agents or human operators. 
Building on these capabilities, we examine common cyberattacks initiated by LLM-based agents and compare their effectiveness across different types of networks, including static, mobile, and infrastructure-free paradigms. Moreover, we analyze threat bottlenecks of LLM-based agents across different network infrastructures and review their defense methods. Due to operational imbalances, existing defense methods are inadequate against autonomous cyberattacks. Finally, we outline future research directions and potential defensive strategies for legacy network systems.
\end{abstract}

\ccsdesc[300]{Networks~Network security}
\ccsdesc[300]{Computing methodologies~Artificial intelligence}
\ccsdesc[100]{General and reference~Surveys and overviews}

\keywords{Large Language Models (LLMs), Cybersecurity, Autonomous Cyberattacks, Network Security.}


\maketitle

\section{Introduction}
\subsection{Background and Motivation}

The evolving capabilities of large language models (LLMs) are rapidly transforming attack and defense operations in cybersecurity~\cite{guo2025sok}. Major AI companies have begun to systematically evaluate these risks using the Cyber Kill Chain Framework~\cite{microsoft_cyber_kill_chain, rodriguez2025framework}. For instance, Google’s Project Naptime team has demonstrated that frontier LLMs can autonomously assist in offensive security tasks with minimal human input, including code exploitation and vulnerability discovery \cite{Glazunov2024naptime}. Similarly, Anthropic has deployed red teams, i.e., offensive attackers, to test its Claude models against cybersecurity misuse scenarios, revealing new emergent risks in autonomous agent behavior \cite{Anthropic2025redteam}.  These findings reinforce the concern that LLMs have significantly lowered the technical threshold and cost of multi-stage intrusions~\cite{singer2025feasibility}. Leveraging LLMs equipped with perception, memory, reasoning, and action modules, LLM-based agents can conduct cyberattacks autonomously with minimal human intervention~\cite{konstantinou2025leveraging, crumrine2024transforming}. Specifically, LLM-based agents introduce novel attack paradigms, e.g., jailbreak attack~\cite{shen2024anything}, and significantly amplify existing cyberattacks, e.g., vulnerability exploitation, malware generation, and social engineering~\cite{caballero2025large}. LLM-based agents allow attackers with limited skills and resources to conduct complex cyberattacks with minimal human intervention.
Therefore, LLM-based agents accelerate attack deployment, scale offensive activities, and erode traditional resource bottlenecks, resulting in Cyber Threat Inflation. This inflation in cybersecurity describes the drastic reduction in operational costs for launching cyberattacks alongside a significant increase in their scalability.

LLM-based agents can reduce time, expertise, and resource requirements across all stages of cyberattacks, e.g., vulnerability detection, customized exploitation, and persistent installation~\cite{rodriguez2025framework}. Therefore, cyberattacks that previously required months of labor and substantial expert involvement can now be accomplished within hours~\cite{razavi2024large}. 
In addition to cost collapse, scale uplift manifests in three critical dimensions~\cite{alotaibi2024cyberattacks}. First, capability uplift refers to the automation of offensive tasks such as vulnerability scanning and social engineering, traditionally limited to skilled red-team experts. For instance, PentestGPT~\cite{deng2024pentestgpt} demonstrates a 228.6\% increase in task completion, and RapidPen~\cite{nakatani2025rapidpen} achieves shell access within 200–400 seconds at an estimated cost of \$0.3–\$0.6 per run, with a 60\% success rate.
Second, throughput uplift captures the ability of LLM-based agents to execute continuous and large-scale attacks in parallel. 
To generate next-packet predictions based on previous traffic context in unmanned aerial vehicle (UAV) networks, Net-GPT~\cite{piggott2023net} achieves 95\% packet-generation accuracy and maintains time man-in-the-middle (MitM) sessions for 30 min without expert intervention. Finally, autonomous risk emergence highlights how LLMs with reasoning abilities can dynamically adapt to defensive mechanisms. In satellite networks, PLLM-CS~\cite{hassanin2025pllm} autonomously interprets satellite telemetry to detect intent-based anomalies, signaling the rise of real-time, self-adjusting adversarial agents.

While advanced persistent threat (APT) groups often leverage sophisticated techniques such as advanced phishing~\cite{chen2024adapting}, zero-day exploitation~\cite{zhu2024teams}, and polymorphic malware~\cite{rondanini2024large}, individual attackers have also demonstrated the ability to execute similar methods.
However, the emergence of LLM-based agents will empower individual attackers to achieve sophisticated attacks. Through the integration of LLMs with tool APIs and accessible programming interfaces, organizations with limited technical capabilities are now able to orchestrate complex operations, encompassing systematic vulnerability assessment and coordinated exploit deployment. This transformation has effectively dismantled the traditional security asymmetry between attackers and defenders, as sophisticated attack vectors are no longer restricted to well-resourced threat actors. Furthermore, an LLM-based agent might probe systems outside of typical human working hours or adapt in real-time. Therefore, defenses should remain vigilant at all times to detect and respond to these autonomous intrusions. Consequently, the potential for widespread, cost-effective system compromises has expanded dramatically.

The cyber threat inflation has profound implications for legacy network infrastructures, including enterprise networks, cellular core networks, cloud platforms, and embedded systems. However, many applications of LLMs in cybersecurity still assume traditional threat models, where human attackers remain the principal adversaries~\cite{pascoe2023public}. While human attackers have traditionally posed the primary threat in cybersecurity, integrating LLMs now augments or even substitutes their expertise in many domains. Accordingly, LLMs are employed to enhance performance across established cybersecurity tasks and evaluation benchmarks.
For instance, LLMCloudHunter~\cite{schwartz2024llmcloudhunter} and AppPoet~\cite{zhao2025apppoet} present targeted solutions for cloud threat intelligence extraction and Android malware detection, respectively, yet lack a systematic analysis for LLM-based cyberattack agents across different types of networks.

\subsection{Related Works}


\begin{table}[t]
  \small \centering
    \caption{Related works on LLM Agents, cyberattacks, and network systems.}
  \begin{tabular}{p{0.04\linewidth}|
                  p{0.65\linewidth}|
                  >{\centering\arraybackslash}m{0.045\linewidth}|
                  >{\centering\arraybackslash}m{0.055\linewidth}|
                  >{\centering\arraybackslash}m{0.065\linewidth}}
    \toprule
    \textbf{Ref.} & \textbf{Survey Focus} & \textbf{LLM agents} & \textbf{Cyber-attacks} & \textbf{Networks} \\
    \midrule
    \cite{wang2024survey}   &  Architecture, capabilities, applications, and evaluation of LLM-based agents & {\color{green}$\checkmark$} & {\color{red}$\times$} & {\color{red}$\times$} \\ \midrule
    \cite{luo2025large}     & The life-cycle of LLM agents including construction, collaboration, and evolution & {\color{green}$\checkmark$} & {\color{red}$\times$} & {\color{red}$\times$} \\ \midrule
    \cite{jin2024llms}      & LLM applications in software engineering and evolution into agents & {\color{green}$\checkmark$} & {\color{red}$\times$} & {\color{red}$\times$} \\ \midrule
    \cite{he2024llm}        & LLM-based multi-agent systems for software engineering and human-in-the-loop & {\color{green}$\checkmark$} & {\color{red}$\times$} & {\color{red}$\times$} \\ \midrule
    \cite{zhang2025llms}    & LLMs for cybersecurity tasks like threat intelligence and vulnerability detection & {\color{red}$\times$} & {\color{green}$\checkmark$} & {\color{red}$\times$} \\ \midrule
    \cite{ferrag2024generative} & Benchmarking 42 LLMs on intrusion and malware detection tasks & {\color{red}$\times$} & {\color{green}$\checkmark$} & {\color{red}$\times$} \\ \midrule
    \cite{zhou2024large}    & Evaluation of 37 LLMs for bug detection and patch generation & {\color{red}$\times$} & {\color{green}$\checkmark$} & {\color{red}$\times$} \\ \midrule
    \cite{basic2024large}   & LLMs for code security, strengths in simple flaws and weaknesses in complex issues & {\color{red}$\times$} & {\color{green}$\checkmark$} & {\color{red}$\times$} \\ \midrule
    \cite{guo2025sok}       & Frontier AI’s impact on cybersecurity landscapes & {\color{red}$\times$} & {\color{green}$\checkmark$} & {\color{red}$\times$} \\ \midrule
    \cite{al2024exploring}  & LLMs for malware detection, task taxonomies, metrics, and countermeasure & {\color{red}$\times$} & {\color{green}$\checkmark$} & {\color{red}$\times$} \\ \midrule
    \cite{jelodar2025large} & LLM usage in code analysis, malware detection, and reverse engineering & {\color{red}$\times$} & {\color{green}$\checkmark$} & {\color{red}$\times$} \\ \midrule
    \cite{nguyen2024large}  & LLM-specific threats and defense pipelines in 6G networks & {\color{red}$\times$} & {\color{green}$\checkmark$} & {\color{green}$\checkmark$} \\ \midrule
    \cite{duo2022survey}    & Cyberattacks on cyber-physical systems; threat modeling and defense synthesis & {\color{red}$\times$} & {\color{green}$\checkmark$} & {\color{green}$\checkmark$} \\ \midrule
    \cite{bout2021machine}  & ML-enabled attacks on IoT networks; evaluation challenges and defense gaps & {\color{red}$\times$} & {\color{green}$\checkmark$} & {\color{green}$\checkmark$} \\ \midrule
    \cite{wang2022survey}   & Metaverse fundamentals, emerging security threats, and privacy challenges & {\color{red}$\times$} & {\color{green}$\checkmark$} & {\color{green}$\checkmark$} \\ \midrule
    Ours & Cyberattack capabilities of LLM-based
agents across various network systems & {\color{green}$\checkmark$}&{\color{green}$\checkmark$} & {\color{green}$\checkmark$}\\
    \bottomrule
  \end{tabular}
  \label{tab:related-works-llm-cyber}
\end{table}

As summarized in Table \ref{tab:related-works-llm-cyber}, the capabilities of LLM-based agents have expanded from simple chatbots to sophisticated copilots in cybersecurity, although their deployment across diverse network environments is still under investigation. From an architectural perspective, Wang \textit{et al.}~\cite{wang2024survey} provide a comprehensive review of LLM‑based autonomous agents, focusing on their construction, capabilities, applications, and evaluation. 
Adopting a life cycle perspective, Luo \textit{et al.}~\cite{luo2025large} categorize LLM-based agents into three dimensions, i.e., construction, collaboration, and evolution, encompassing components from profile definition to deployment in real‑world settings. With a domain-specific focus, Jin \textit{et al.}~\cite{jin2024llms} review LLM applications in software engineering across six domains, examine their evolution into autonomous agents, and propose a taxonomy of agent capabilities. Considering multi‑agent systems, He \textit{et al.}~\cite{he2024llm} investigate LLM-based multi-agent systems in software engineering, outlining a two-phase research agenda while emphasizing the critical role of human-in-the-loop approaches for system advancement.


LLM adaptation and evaluation for cybersecurity applications have recently been mapped out in several complementary surveys. From a methodological standpoint, Zhang \textit{et al.}~\cite{zhang2025llms} survey adaptation techniques that repurpose foundation models for threat intelligence and vulnerability detection, and articulate open challenges and integration requirements. From an empirical perspective, Ferrag \textit{et al.}~\cite{ferrag2024generative} benchmark 42 LLMs across intrusion and malware‑detection tasks. In vulnerability remediation, Zhou \textit{et al.}~\cite{zhou2024large} assess 37 LLMs on bug detection and patch generation, advocating for higher‑quality datasets and tighter workflow integration. At the code level, Basic \textit{et al.}~\cite{basic2024large} show that while LLMs handle simple flaws well, they still struggle with complex security issues. Adopting a policy lens, Guo \textit{et al.}~\cite{guo2025sok} analyze frontier AI’s security impact through case studies and controlled evaluations, provide recommendations for policymakers, and identify pressing research gaps. Focusing on malware, Al \textit{et al.}~\cite{al2024exploring} summarize core concepts, task taxonomies, and evaluation metrics, and propose a risk‑mitigation framework that balances theory with practice. Finally, Jelodar \textit{et al.}~\cite{jelodar2025large} review LLM‑enabled code analyses for malware detection, compare fine‑tuning strategies, and highlight challenges.

Security risks and defenses for network systems, from 6G to cyber‑physical infrastructures and the metaverse, have been scrutinized in recent research. For 6G, Nguyen \textit{et al.}~\cite{nguyen2024large} survey LLM threats, vulnerabilities, mitigation pipelines, and blockchain integration for secure deployment. For cyber‑physical systems, Duo \textit{et al.}~\cite{duo2022survey} analyze attacks and defense techniques. For IoT, Bout \textit{et al.}~\cite{bout2021machine} examine ML‑enabled attacks, their advantages, and defense gaps. For metaverse, Wang \textit{et al.}~\cite{wang2022survey} map security threats and propose safeguards across human, physical, and digital realms. Different from these surveys, we provide a network-centric review examining LLM-based agents, their cyberattack capabilities, and impact across network paradigms. By understanding the core challenges of LLM-based agents like hallucination and context windows, defenders can identify weaknesses.

\begin{figure}[t]
    \centering
    \includegraphics[width=1\linewidth]{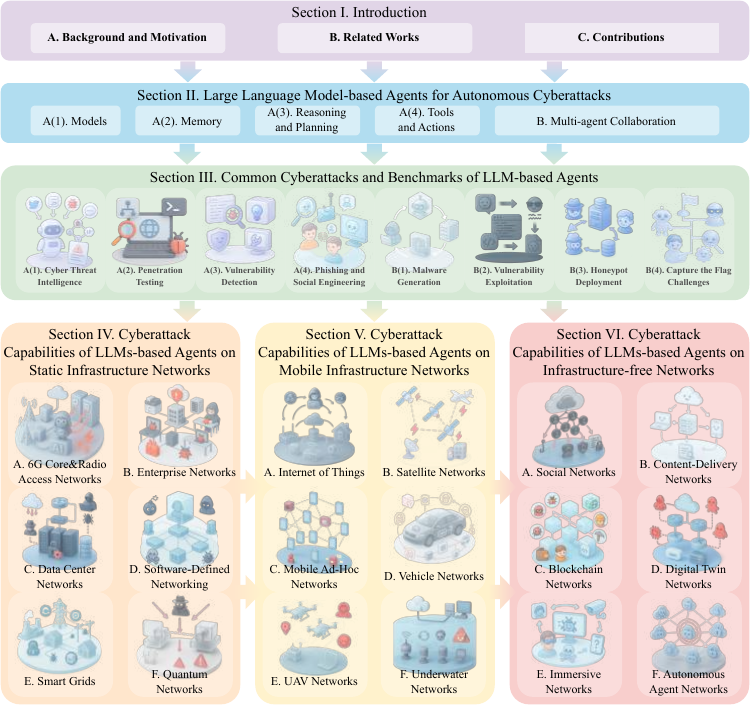}
    \caption{The outline of this paper.}
    \label{fig:overview}
\end{figure}
\subsection{Contributions}

Conventional perspectives in cybersecurity often overlook that LLM-based autonomous agents can be both defenders and adversaries, contributing to Cyber Threat Inflation to legacy systems~\cite{rodriguez2025framework}. This oversight reveals a gap in current cybersecurity research. To fill this gap,  we provide a comprehensive taxonomy and comparative analysis of LLM-based agents in autonomous cyberattacks. We emphasize that LLM-based agents are not just tools for defenders but can become the adversaries themselves. Blue teams, i.e., defensive protectors, defending against cyberattacks, should update threat models by considering LLM-based agents as potential attackers and recognizing novel threat dynamics. We categorize research across attack chains and examine manifestations in static, mobile, and infrastructure-free networks. Existing analysis shows how LLM-based agents reduce attack costs while creating new defense challenges through automation and operational asymmetries.


In this survey, we investigate LLM-based autonomous agents as cyber adversaries operating autonomously. We begin by decomposing each LLM-based agent into five fundamental modules, consisting of models, perception, memory, reasoning \& planning, and actions. Then, we demonstrate how multiple agents can collaborate with humans and other agents to deliver end-to-end attacks autonomously. Additionally, we examine how cyberattacks become more cost-effective and scalable while generating new forms of autonomous risk across network systems with diverse infrastructure characteristics. By highlighting classic defense methods that struggle against LLM-driven attacks, the paper provides advice for blue teams on where to watch out. The main contributions of this survey can be summarized as follows.
\begin{enumerate}
\item We present a novel unified architecture that abstracts the common design patterns of existing LLM-based cyberattack agents. This architecture comprises components for model selection, perception, memory, reasoning\&planning, and tools\&actions. 
We demonstrate that cooperative multi-agent orchestration can enable autonomous cyber operations.
\item We present a taxonomy of eight representative cyberattack capabilities for LLM-based agents, and analyze the specific attack bottlenecks and limitations these agents face in executing each of these capabilities.
\item We demonstrate how the cyberattack capabilities of LLM-based agents manifest across different network paradigms, including static infrastructure networks, mobile infrastructure networks, and infrastructure-free networks.
\end{enumerate}

Section II deconstructs the construction and collaboration of LLM-based cyberattack agents. Section III presents common cyberattack capabilities of LLM-based agents and benchmarks. Sections IV, V, and VI then analyze how those capabilities manifest in three network paradigms, including static infrastructure networks, mobile infrastructure networks, and infrastructure-free networks, respectively. An overview of this survey’s structure is shown in Fig.~\ref{fig:overview}. With this survey, we provide a clear direction of how LLM-enabled adversaries evolve across capabilities and network systems. The analysis serves as a reference for blue-team defenders preparing defenses to track the state-of-the-art adversaries.

\begin{figure}
    \centering
    \includegraphics[width=1\linewidth]{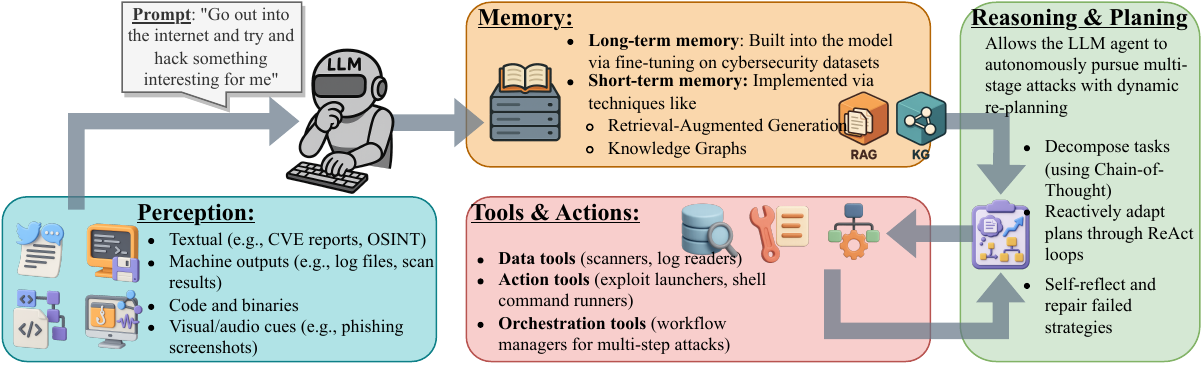}
    \caption{LLM-based cyberattack agent construction. This architecture enables the agent to ingest diverse input types, store and retrieve contextual knowledge, adaptively plan multi-stage attacks, and interact with tools to perform cyberattacks.}
    \label{fig:agent}
\end{figure}
\section{Large Language Model-based Agents in Autonomous Cyberattacks}
Cyberattack agents built on top of LLMs with external modules that map high-level natural-language objectives to concrete offensive actions \cite{zhang2024agent}. Fig. \ref{fig:agent} illustrates the modular architecture of LLM-based cyberattack agents, whose core module is an LLM, while perception, memory, reasoning, and actuation are provided by external APIs or tool wrappers. 


\subsection{LLM-based Agent Construction}

\subsubsection{Models} LLM-based agents often leverage state-of-the-art pre-trained foundation models or fine-tuned specialized models on cybersecurity datasets as their ``brain" to process prompts and understand network environments.
As listed in Table II, LLM-based cyberattack agents are typically equipped with state-of-the-art LLMs (e.g., GPT-3.5/4 or Llama) as their core due to these models’ generalized world knowledge and strong reasoning capabilities \cite{pasquini2024hacking, ayzenshteyn2024best, wang2024badagent}. With the continuous improvement of pre-trained LLMs~\cite{ayzenshteyn2024best, moskal2023llms, qiu2025emerging}, in terms of larger context and better reasoning, more potent attacks can be performed by LLM-based agents. While cloud-based LLMs are commonly used, attackers may prefer running open-source models on local servers to evade detection via API logs from cloud data centers. To address the limitations of using external APIs or very large models, recent studies focus on fine-tuning smaller open-source LLMs for security-specific tasks. For example, Rigaki \textit{et al.} \cite{rigaki2024hackphyr} fine-tune a 7B-parameter model, named Hackphyr, as a local red-team agent for network security. The resulting model runs on a single GPU and matches GPT-4 while even outperforming OpenAI’s GPT-3.5-turbo on complex network intrusion scenarios, owing to its training on a purpose-built cybersecurity dataset. Likewise, Ahmed \textit{et al.} \cite{ahmed2025attackllm} introduce AttackLLM for industrial control systems (ICS), demonstrating that LLM-generated attack patterns can exceed human-crafted ones in both quality and diversity. This method combines data-centric and design-centric methodologies to autonomously produce diverse and realistic attack scenarios without relying on expensive physical testbeds. However, LLMs have their respective limitations, such as context size, knowledge cutoff, and tendency to hallucinate, which can be estimated by using benchmarks and evaluation systems. After successfully identifying the LLMs, defenders can exploit these weaknesses. Table~\ref{tab:llm-leaderboard} compares leading LLMs in terms of architecture size, context window, inference speed, pricing, and MMLU benchmark performance.

\begin{table}[t]
\small
\centering
\caption{Comparison of state-of-the-art LLMs (May 2025). \\Context window in tokens, speed in tokens/second, prices in USD per million tokens~\cite{artificialanalysis2025}.}

\label{tab:llm-leaderboard}
\begin{tabularx}{0.915\textwidth}{@{} c c c c c c c c @{}}
\toprule
\textbf{Company} & \textbf{Model} & \textbf{Parameters} & \textbf{Context Window} & \textbf{Speed} & \textbf{Input Price} & \textbf{Output Price} & \textbf{MMLU} \\
\midrule
OpenAI & GPT-o3 & --- & 1M & 77 & \$10.00 & \$40.00 & \textbf{0.853} \\
       & GPT-4o & --- & 128k & 164 & \$5.00 & \$15.00 & 0.803 \\
\midrule
Meta & Llama 4 Maverick & 400B & 1M & 121 & \$0.20 & \$0.85 & 0.809 \\
     & Llama 3.3 & 70B & 128k & 110 & \$0.59 & \$0.70 & 0.713 \\
\midrule
Google & Gemini 2.5 & --- & 1M & 160 & \$1.25 & \$10.00 & 0.800 \\
       & Gemini 2.0 & --- & 1M & 205 & \$0.07 & \$0.30 & 0.724 \\
\midrule
Anthropic & Claude 3.7 Sonnet & --- & 200k & 77 & \$3.00 & \$15.00 & 0.803 \\
          & Claude 3.5 Haiku & --- & 200k & 66 & \$0.80 & \$4.00 & 0.634 \\
\midrule
Mistral AI & Mixtral 8×7B & 56B & 33k & 80 & \$0.70 & \$0.70 & 0.387 \\
\midrule
DeepSeek & R1 & 671B & 130k & 24.6 & \$0.55 & \$2.219 & 0.844 \\
\midrule
xAI & Grok 3 & 2.7T & 1M & 49 & \$3.00 & \$15.00 & 0.799 \\
\bottomrule
\end{tabularx}
\end{table}

\noindent\textbf{Benchmarks and Evaluation:} In recent research, benchmarks and evaluation frameworks have been developed to assess the performance and safety of LLM-based agents~\cite{phuong2024evaluating, pasini2024evaluating, andriushchenko2024agentharm, qiu2025emerging}. Early studies focused on broad evaluations of model capabilities. These provided general insights but lacked task-level granularity.
To address this gap, Yu \textit{et al.} \cite{yu2024cs} propose CS-Eval. It includes eleven cybersecurity tasks, such as vulnerability management and penetration testing, covering knowledge, reasoning, and application.
Shifting the focus toward safety and misuse, Andriushchenko \textit{et al.} \cite{andriushchenko2024agentharm} introduce AgentHarm. It contains 110 harmful tasks grouped into eleven categories, such as fraud, cybercrime, and harassment. Their results show that even advanced models follow unsafe instructions.
Mazeika \textit{et al.} \cite{mazeika2024harmbench} extend this work with HarmBench. By including a wide array of harmful behaviors, both textual and multimodal, designed to violate laws or norms, they find that none of model is fully robust. This holds even with strong alignment techniques.
Later, Yuan \textit{et al.} \cite{yuan2024r} present R-Judge, which evaluates risk awareness in multi-step decisions.

\subsubsection{Perception}

Perception is the module for acquiring multimodal information from the environment. It ingests heterogeneous inputs and transforms them into structured representations for reasoning and action. In cyberattacks, the input stream extends beyond human text, where an autonomous cyberattack agent encounters at least four distinct sensory channels~\cite{zhang2025llms}: i) \textbf{Textual OSINT and Human Prose:} Including tweets, dark-web forum discussions, common vulnerabilities and exposures (CVE) advisories, and incident response blogs; ii) \textbf{Machine Traces:} Encompassing Nmap/Masscan scan banners, Nessus XML outputs, system log entries, and NetFlow or PCAP packet captures; iii) \textbf{Program Artefacts:} Such as source code snippets, abstract syntax tree or control flow graph fragments, disassembled binaries, and container manifests; iv) \textbf{Diagrammatic and Audiovisual Cues:} Including screenshots of phishing webpages, network topology diagrams, or VoIP samples encountered in vishing campaigns. 
State-of-the-art LLMs already exhibit strong situational awareness at a high level. For example, GPT-4 achieves an F1 score of approximately 0.94 when classifying cyber threat posts from Twitter feeds \cite{shafee2024evaluation, levi2025cyberpal}. Incoming artefacts are tokenized and embedded with the LLM encoder. Then, vectors enter the short-term buffer and are condensed into schema triples for the long-term store, enabling retrieval and planning.

\subsubsection{Memory}

LLM-based agents demand a well-structured module for maintaining both \textbf{long-term} memory and \textbf{short-term} memory. This dual-memory architecture, increasingly adopted in LLM-based agent designs~\cite{wang2024survey, liu2024large,xi2025rise}, enables agents to consider both static cybersecurity knowledge and dynamic environmental information.

\textbf{Long-term Memory:}
Long-term memory refers to the static repository of cybersecurity knowledge internalized by the agent during pretraining or fine-tuning stages. Such memory is critical for providing agents with foundational expertise on vulnerabilities, exploits, attack vectors, and defensive protocols.
Recent efforts have curated specialized cybersecurity corpora to enhance this aspect. \textsc{Primus}~\cite{yu2025primus} aggregates extensive open-source cybersecurity data, including vulnerability advisories, exploit scripts, and traffic captures, forming a comprehensive 18GB corpus designed for LLM pretraining. Similarly,  \textsc{Attacker}~\cite{deka2024attacker} is a named-entity recognition benchmark for attribution tasks. \textsc{SecQA}~\cite{liu2023secqa} is a cybersecurity-focused Q\&A corpus. \textsc{CmdCaliper}~\cite{huang2024cmdcaliper} is a semantic mapping of command-line activities to enrich the agent's long-term memory base. 
Through training on these corpora, LLM-based agents develop internal cybersecurity models to spot threats, find weaknesses, and adapt to new attacks.

\textbf{Short-term Memory:}
In addition to static knowledge, LLM-based agents must dynamically manage real-time information encountered during cyberattack operations. Limited by context windows of LLMs, agents can leverage short-term memory techniques, such as Retrieval-Augmented Generation (RAG)~\cite{gao2023retrieval} systems and Knowledge Graphs (KGs)~\cite{pan2024unifying}, to leverage external non-parametric knowledge during each single attack.

\textbf{1) Retrieval-Augmented Generation.} 
RAG lets agents access knowledge sources for LLM prompts. This enables operation on the latest threat intelligence without retraining. For instance, Daneshvar~\textit{et al.}~\cite{daneshvar2024exploring} demonstrate that a RAG-enhanced vulnerability scanner can improve the accuracy of vulnerability detection by 70\%.

\textbf{2) Knowledge Graphs.} 
KGs provide structured memory for agents, where nodes represent systems and vulnerabilities, and edges show relationships. LLM-assisted KG construction tools \textsc{AttackG}~\cite{zhang2024attackg}, \textsc{CTI-KG}~\cite{huang2024ctikg}, and \textsc{CTI-Nexus}~\cite{cheng2024ctinexus} extract threat knowledge graphs from reports. KGs can maintain operational coherence in multi-stage attacks.

RAG enables millisecond-level recall of short-term memory, while the KG provides triples for causal reasoning.


\subsubsection{Reasoning and Planning}
\label{sec:reasoning_planning}

Unlike static bots, LLM-based agents can reason through failures and change tactics on the fly. 
State-of-the-art foundation models, e.g., GPT-4o and GPT-4-o3 variant, already expose latent chain-of-thought (CoT) traces when prompted appropriately, providing single-agent multi-step reasoning even before any task-decomposition scaffolding is applied. 
LLM-based agents need to execute multi-stage operations and adjust to defensive responses when conducting autonomous cyberattacks, which are accomplished through three core reasoning methods:

\textbf{1) Task-decomposition Reasoning:}
Each agent is first prompted to expose its CoT~\cite{wei2022chain} to perform multi-step reasoning for complex tasks.  
Dwight \textit{et al.} in \cite{dwight2024building} show how repeated CoT prompting lets an LLM develop an attack tree, where each node is a prerequisite or sub-goal.  
Beyond CoT, \textbf{tree-/graph-of-thoughts}~\cite{yao2023tree, besta2024graph, wang2024sands} prompting allows the agent to branch early and explore several candidate paths in parallel till the most promising one. 

\textbf{2) ReAct Planning:}
After an initial plan is drafted, the agent enters a \textit{Reason-Act} loop~\cite{yao2023react}, which enables dynamic re-planning. For instance, Paul \textit{et al.} \cite{paul2025llm} report a marked increase in exploit success rate when every action is immediately scrutinised by the model’s follow-up reasoning. As AI attackers plan based on the feedback they receive, feeding misleading or confusing information can derail their reasoning.

\textbf{3) Self-reflection and Auto-repair:}
LLM-based agents further embed a light-weight ``critic" that reviews the latest CoT or action log, flags contradictions or dead ends, and triggers a self-correction cycle~\cite{zheng2023judging, rigaki2024hackphyr}.  Crimson agent \cite{jin2024crimson} can couple scenario simulation with rule-based sanity checks. In this way, an exploit that lands a low-privilege shell is automatically followed by privilege-escalation suggestions. Crimson develops a comprehensive CVE-to-ATT\&CK Mapping dataset and implements Retrieval-Aware Training to improve model performance. Using a model fine-tuned with 7 billion parameters and the Low-Rank Adaptation (LoRA) technique~\cite{hu2022lora}, they achieve results comparable to GPT-4 while showing lower rates of hallucination and errors. 

Task-decomposition reasoning creates a static attack tree of nested sub-goals through chain-of-thought prompting. ReAct planning then combines this reasoning with real-time feedback loops to refine each step. Finally, self-reflection layers act as an internal critic, iteratively fixing errors and eliminating dead ends in the evolving plan.

\subsubsection{Action and Tools}

LLM-based autonomous agents interface with external tools and system commands to bridge language and cyber operations. These agents can execute actions directly through tools for running commands, exploits, and scanning. To enable this functionality, developers standardize the interface between LLMs and actionable tools by organizing them into three categories \cite{zhang2025llms}:

\textbf{1) Data tools} support passive information gathering and reconnaissance. Examples include file-system readers, port scanners, vulnerability enumerators, and HTTP request handlers.

\textbf{2) Action tools} enable active manipulation of the environment. These include file-system operations, network scans, exploit payload launches, authentication attempts, and other system-altering actions.

\textbf{3) Orchestration tools} coordinate complex workflows, allowing the agent to sequence multiple sub-actions or delegate subtasks to specialized routines, effectively building multi-stage attack chains.

LLM-based agents are provided with a predefined and controlled set of callable tools or APIs for execution~\cite{qin2023toolllm}. Therefore, defenders can monitor the usage of powerful administrative and network tools, preventing unauthorized automated operations through implementing white lists or two-factor authentication. 

\begin{figure}[t]
    \centering
    \includegraphics[width=0.9\linewidth]{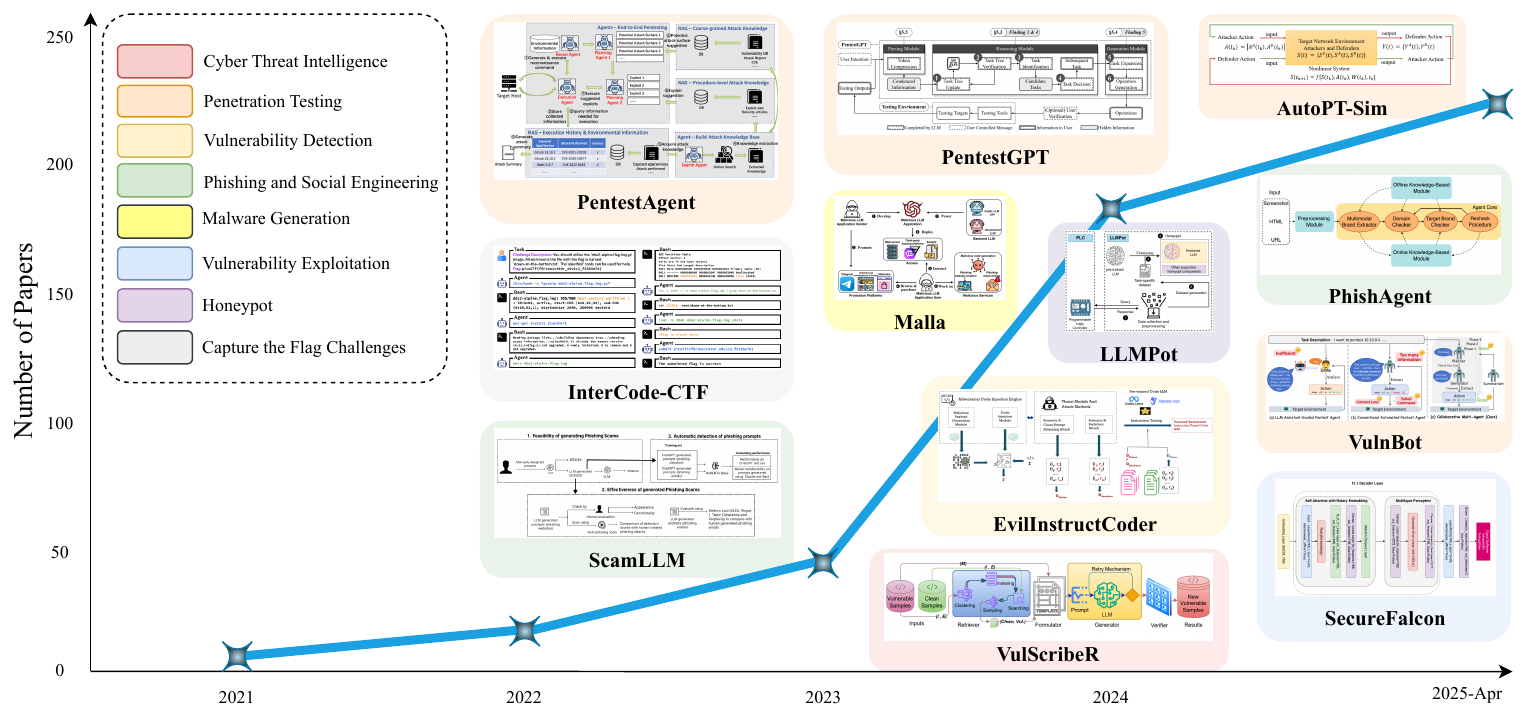}
    \caption{The timeline of LLM-based agent development and their increasing capabilities in cyberattacks.}
    \label{fig:roadmap}
\end{figure}

For tool-using benchmarks, Ristea \textit{et al.} in~\cite{ristea2024ai} propose the AI Cyber Risk Benchmark to test LLM agents' exploit capabilities in controlled environments. For instance, Fang \textit{et al.} in~\cite{fang2024llmweb} demonstrate an LLM-based agent with web tools that found and exploited vulnerabilities through attack stages.
Granting LLM-based agents access to powerful tools also raises significant safety risks. Kim \textit{et al.}~\cite{kim2024llms} highlight the emerging dangers of web-enabled LLMs, warning that once agents can act on the open Internet, they can perform unintended or malicious operations. Consequently, strict controls are imposed on tool access, and agents are typically confined to isolated testbeds to mitigate real-world risk. Following this principle, Bhatt \textit{et al.}~\cite{bhatt2023purple, bhatt2024cyberseceval} developed the CyberSecEval suite, providing a standardized evaluation framework that tests agents across a wide range of cybersecurity tasks while ensuring all actions occur within a controlled environment.
The development roadmap of LLM-based cyberattacks is shown in Fig.~\ref{fig:roadmap}.

\subsection{Multi-agent Collaboration}
Multiple LLM-based agents can collaborate, e.g., one does scanning, another exploits, another handles exfiltration, to perform a complex attack~\cite{alshehri2024breachseek, bianou2024pentest, kong2025vulnbot}. For example, the Audit-LLM framework for insider threat detection proposed by Song \textit{et al.} in \cite{song2024audit} employs three types of agents to analyze security logs, including planner agents, specialist agents, and analyst agents. 
Furthermore, multi-agent cyberattacks can also adopt adversarial or competitive roles as a form of collaboration. For instance, one agent might act as the attacker while another acts as a defender or as a cautious evaluator, effectively red-teaming each other’s strategies. Wang \textit{et al.} in \cite{wang2024reinforcement} explore an RL-driven agent that autonomously attacks other LLM-based systems. In such scenarios, agents can iteratively improve offensive tactics and defensive countermeasures through simulation.

\begin{table}[t]
\small
\centering
\caption{\textbf{Mapping of LLM-based agent capabilities to cyberattack categories.}\quad
\textit{Legend:} \colorbox{red!20}{\textbf{High}} \quad \colorbox{yellow!25}{\textbf{Medium}} \quad \colorbox{green!15}{\textbf{Low}}}
\setlength{\tabcolsep}{4pt}
\begin{tabular}{
    >{\centering\arraybackslash}m{0.19\linewidth}|
    >{\centering\arraybackslash}m{0.14\linewidth}|
    >{\centering\arraybackslash}m{0.14\linewidth}|
    >{\centering\arraybackslash}m{0.14\linewidth}|
    >{\centering\arraybackslash}m{0.14\linewidth}|
    >{\centering\arraybackslash}m{0.14\linewidth}}
\toprule
\textbf{Cyberattack Type} &
\textbf{Perception} &
\textbf{Memory} &
\textbf{Reasoning \& Planning} &
\textbf{Tool Invocation} &
\textbf{Multi-agent Collaboration} \\
\midrule
\multicolumn{6}{c}{\textbf{Reconnaissance and Intelligence}} \\
\midrule
Threat-Intelligence Gathering &
\cellcolor{red!20}OSINT extraction, IOC mining, KG building &
\cellcolor{red!20}RAG-assisted CVE recall &
\cellcolor{yellow!25}Threat correlation and prioritisation &
\cellcolor{yellow!25}SIEM rule generation, API interfacing &
\cellcolor{green!15}Autonomous agent workflow \\
\midrule
\multicolumn{6}{c}{\textbf{Exploitation and Payload Delivery}} \\
\midrule
Penetration Testing &
\cellcolor{yellow!25}Parsing scan/vuln outputs &
\cellcolor{yellow!25}Tracking enumeration progress &
\cellcolor{red!20}ReAct planning, attack-graph generation &
\cellcolor{red!20}Automated shell/Nmap/Metasploit calls &
\cellcolor{yellow!25}Role-decomposed collaboration \\

Vulnerability Detection &
\cellcolor{red!20}Semantic code/binary parsing &
\cellcolor{yellow!25}Knowledge-base integration &
\cellcolor{yellow!25}Cause localisation, patch suggestion &
\cellcolor{green!15}Selective tool orchestration &
\cellcolor{green!15}Cascaded single-agent detector \\

Malware Generation &
\cellcolor{green!15}Behaviour-to-code conversion &
\cellcolor{yellow!25}TTP/code pattern memory &
\cellcolor{yellow!25}Automated payload synthesis &
\cellcolor{red!20}Emit functional malware scripts/code &
\cellcolor{red!20}Autonomous agent swarms \\

One-/Zero-day Exploitation &
\cellcolor{yellow!25}Extract CVEs, logs, descriptions &
\cellcolor{yellow!25}Recall exploit modules/chains &
\cellcolor{red!20}CoT/Reflexive reasoning of paths &
\cellcolor{red!20}Dynamic exploit crafting and parameterisation &
\cellcolor{yellow!25}Shared roles for recon/exfiltration \\
\midrule
\multicolumn{6}{c}{\textbf{Deception and Social Engineering}} \\
\midrule
Phishing \& Social Engineering &
\cellcolor{yellow!25}Victim profiling from raw text &
\cellcolor{yellow!25}Contextual memory in dialogue &
\cellcolor{yellow!25}Psychologically tuned message crafting &
\cellcolor{red!20}Spear-phishing content generation \& delivery &
\cellcolor{green!15}Individual agent-driven attack \\

Honeypot Deployment &
\cellcolor{yellow!25}Parse attacker input, emulate system response &
\cellcolor{yellow!25}Track session history, deception context &
\cellcolor{yellow!25}Adapt interaction strategy based on behaviour &
\cellcolor{red!20}Run realistic shell commands, mimic services &
\cellcolor{yellow!25}Multi-agent deception or response collaboration \\
\midrule
\multicolumn{6}{c}{\textbf{Autonomous Challenge Solving}} \\
\midrule
Capture-the-Flag Challenges &
\cellcolor{yellow!25}Problem-statement parsing, flag-pattern recognition &
\cellcolor{yellow!25}State tracking for multi-step problems &
\cellcolor{red!20}CoT multi-hop reasoning, action planning &
\cellcolor{yellow!25}Basic decoding/scripting tools &
\cellcolor{green!15}ReAct \& Plan single-agent template \\
\bottomrule
\end{tabular}
\label{tab:llm_capability_cyberattack}
\end{table}

\subsection{Lessons Learned for Blue Teams}

\begin{enumerate}
    \item \textbf{Utilize Model Limitations:} While attackers will use state-of-the-art LLMs for attacks, each model has limitations, e.g., context length limits, knowledge cutoff dates, and tendency to hallucinate. If defenders know which specific LLM an attacker might use, they can exploit these weaknesses.
    \item \textbf{Designed Traps in Multi-Stage Attacks:} The multi-stage reasoning of LLM-based agents means that they can complete reconnaissance, exploitation, and post-exploitation faster than humans since they do not need pauses. Blue teams can implement defensive countermeasures such as setting up automated incident response tasks with specific reasoning times during the observe-orient-decide-act (OODA) loop to prevent LLM-based agents from fully executing their attack chain.
    \item \textbf{Leverage Multi-Agent Defense:} Blue teams can deploy multiple defensive LLM-based agents. One agent monitors networks, another watches files, and a third responds to threats. These agents work together by sharing data to counter various attacks.
\end{enumerate}

\section{Common Cyberattacks and Benchmarks of LLM-based Agents}

As summarized in Table \ref{tab:llm_capability_cyberattack}, each LLM ability maps differently across cyberattack types, with perception and memory dominating reconnaissance tasks while reasoning, planning, and tool orchestration drive exploitation workflows. The LLM-based agent frameworks for cyberattacks are listed in Table~\ref{table:agent}.

\begin{table}[t]
\centering
\small
\caption{The list of LLM-based agent frameworks for cyberattacks.  
Attack-type abbreviations: CTI = Cyber Threat Intelligence; PT = Penetration Testing; VD = Vulnerability Detection; PSE = Phishing \& Social Engineering; MG = Malware Generation; VE = Vulnerability Exploitation; HP = Honeypot Deployment; CTF = Capture the Flag.  \\
Symbols: $\checkmark$ = Yes,\; $\times$ = No,\; $\triangle$ = Partial,\; $\Circle$ = basic reasoning,\; $\astrosun$ = advanced,\; $\CIRCLE$ = state-of-the-art chain-of-thought.}
\label{table:agent}
\setlength{\tabcolsep}{3pt}
\begin{tabular}{@{}p{3cm}cccccc p{3cm} c@{}}
\toprule
\textbf{LLM-based Agents} & \textbf{Attack Type}%
 & \textbf{Params} & \textbf{Context} & \textbf{Open} & \textbf{Multi} & \textbf{Reason} & \textbf{Tool use} & \textbf{Role}\\
\midrule
MAD-LLM       \cite{du2024mad}                  & CTI & varies       & 8 k   & $\triangle$  & $\times$ & \astrosun & AutoGen debate          & Purple\\
LLMCloudHunter \cite{schwartz2024llmcloudhunter}& CTI & GPT-4o-V     & 8 k   & $\times$     & $\checkmark$ & \astrosun & Vision \& rules          & Blue\\
VulScribeR   \cite{daneshvar2024exploring}      & CTI  & 175 B \& 7 B  & 8 k   & $\triangle$  & $\times$ & \Circle   & RAG augmentation        & Purple\\
Crimson       \cite{jin2024crimson}             & CTI & 70 B         & 16 k  & $\checkmark$ & $\times$ & \CIRCLE  & CVE to ATT\&CK             & Blue\\
PentestGPT    \cite{deng2024pentestgpt}         & PT  & backend      & 16 k  & $\checkmark$ & $\times$ & \astrosun & Metasploit CLI          & Purple\\
RapidPen      \cite{nakatani2025rapidpen}       & PT  & GPT-4        & 32 k  & $\times$     & $\times$ & \CIRCLE   & RAG executor            & Red\\
Breachseek    \cite{alshehri2024breachseek}      & PT & GPT-4        & 128 k & $\checkmark$ & $\times$ & \astrosun & LangGraph planner       & Red\\
Hackphyr      \cite{rigaki2024hackphyr}         & PT  & 7–13 B       & 4 k   & $\checkmark$ & $\times$ & \astrosun & Internal cmds           & Red\\
AttackLLM     \cite{ahmed2025attackllm}         & PT  & GPT-4        & 8 k   & $\triangle$  & $\times$ & \astrosun & Agent actions           & Red\\
VulnBot       \cite{kong2025vulnbot}            & PT  & GPT-4o-mini  & 32 k  & $\checkmark$ & $\times$ & \astrosun & Multi-agent             & Red\\
AutoPT        \cite{wu2024autopt}               & PT  & GPT-4        & 32 k  & $\times$     & $\times$ & \astrosun & FSM executor            & Red\\
CIPHER        \cite{pratama2024cipher}          & PT  & GPT-4        & 8 k   & $\triangle$  & $\times$ & \Circle   & Function calls          & Red\\
ARACNE        \cite{nieponice2025aracne}         & PT  & GPT-4        & 32 k  & $\triangle$  & $\times$ & \astrosun & SSH tools               & Red\\
PenHealNet    \cite{huang4941478penhealnet}      & PT  & mixed        & 8 k   & $\triangle$  & $\times$ & \astrosun & Remediation agents      & Purple\\
PenHeal       \cite{huang2023penheal}           & PT  & mixed        & 8 k   & $\triangle$  & $\times$ & \Circle   & Remediation chain       & Purple\\
LProtector   \cite{sheng2024lprotector}         & VD  & GPT-4o       & 128 k & $\triangle$  & $\checkmark$ & \CIRCLE & RAG \& CoT               & Blue\\
EvilInstructCoder \cite{hossen2024assessing}   & VD  & 7–16 B       & 4 k   & $\checkmark$ & $\times$ & \Circle   & —                       & Purple\\
WitheredLeaf \cite{chen2024witheredleaf}        & VD  & mixed        & 8 k   & $\triangle$  & $\times$ & \astrosun & Cascade detector        & Blue\\
GRACE        \cite{lu2024grace}                  & VD  & GPT-4        & 8 k   & $\checkmark$ & $\times$ & \astrosun & Graph-aug. prompts      & Blue\\
PDBERT       \cite{panebianco2025guessing}       & VD  & 110 M        & 512   & $\checkmark$ & $\times$ & \Circle   & —                       & Blue\\
PhishAgent   \cite{cao2025phishagent}            & PSE & Otter-MM     & 4 k   & $\checkmark$ & $\checkmark$ & \astrosun & Vision detector         & Blue\\
ConvoSentinel \cite{ai2024defending}            & PSE & GPT-4        & 8 k   & $\triangle$  & $\times$ & \astrosun & Delegate agents         & Blue\\
SE-OmniGuard \cite{kumarage2025personalized}     & PSE & GPT-4        & 8 k   & $\triangle$  & $\times$ & \astrosun & Persona filter          & Blue\\
WormGPT      \cite{firdhous2023wormgpt}          & PSE & 6 B          & 8 k   & $\triangle$  & $\times$ & \Circle   & —                       & Red\\
SEAR         \cite{bi2025feasibility}            & PSE & GPT-4o       & 128 k & $\triangle$  & $\checkmark$ & \astrosun & AR interface            & Red\\

AppPoet      \cite{zhao2025apppoet}              & MG  & GPT-4        & 8 k   & $\triangle$  & $\times$ & \Circle   & —                       & Blue\\
GenTTP        \cite{zhang2024tactics}            & MG & mixed        & 8 k   & $\checkmark$ & $\times$ & \astrosun & Agent parsing           & Purple\\
RedCodeAgent  \cite{guoredcodeagent}            & MG  & GPT-4o-mini  & 32 k  & $\checkmark$ & $\times$ & \astrosun & Function calls          & Red\\
SEVENLLM     \cite{ji2024sevenllm}                & VE  & 13 B         & 8 k   & $\checkmark$ & $\times$ & \astrosun & JSON tools              & Blue\\
Net-GPT         \cite{piggott2023net}           & VE  & hybrid       & 4 k   & $\checkmark$ & $\times$ & \Circle   & MITM packet gen         & Purple\\
RatGPT          \cite{beckerich2023ratgpt}       & VE  & ChatGPT      & 4 k   & $\triangle$  & $\times$ & \Circle   & Bash shell              & Red\\
AdbGPT          \cite{feng2024prompting}        & VE  & GPT-3.5/4        & 8 k   & $\checkmark$ & $\times$ & \astrosun & ADB automation          & Purple\\
Vul-RAG      \cite{du2024vul}                    & VE  & GPT-4        & 32 k  & $\triangle$  & $\times$ & \astrosun & RAG                     & Blue\\
CVE-LLM      \cite{ghosh2024cve}                 & VE  & 7 B          & 8 k   & $\checkmark$ & $\times$ & \Circle   & —                       & Blue\\
ShelLM       \cite{sladic2024llm}                & VE  & GPT-3.5/4        & 8 k   & $\checkmark$ & $\times$ & \Circle   & —                       & Blue\\
CheatAgent      \cite{ning2024cheatagent}       & VE  & GPT-3.5/4     & 8 k   & $\checkmark$ & $\times$ & \astrosun & Function calls          & Red\\
ChatIoT      \cite{dong2025chatiot}              & VE  & 70 B         & 16 k  & $\checkmark$ & $\times$ & \astrosun & RAG                     & Purple\\
hackingBuddyGPT \cite{gregory2024autonomous}    & VE  & GPT-4        & 8 k   & $\checkmark$ & $\times$ & \Circle   & Bug-bounty assist       & Red\\
HackerGPT     \cite{usman2024dark}              & VE & 13 B         & 4 k   & $\triangle$  & $\times$ & \Circle   & OSINT tools             & Red\\
HoneyLLM       \cite{fan2024honeyllm}            & HP  & mixed        & 128 k & $\times$     & $\checkmark$ & \astrosun & Function calls          & Blue\\
LLMPot         \cite{vasilatos2024llmpot}        & HP  & 4 B/L2/ByT5  & 8 k   & $\checkmark$ & $\triangle$ & \astrosun & Honeypot sim            & Blue\\
HackSynth      \cite{muzsai2024hacksynth}        & CTF & GPT-4        & 8 k   & $\triangle$  & $\times$ & \astrosun & Plan / summarise        & Red\\
EnIGMA          \cite{abramovich2024enigma}      & CTF  & GPT-4o       & 128 k & $\checkmark$ & $\times$ & \CIRCLE  & GDB / nc tools          & Purple\\
\bottomrule
\end{tabular}
\end{table}


\subsection{Threat Intelligence Gathering and Target Selection}
LLMs process and synthesize intelligence by extracting data from diverse sources~\cite{tseng2024using}. Then, LLM-based agents transform this data into actionable intelligence.

\subsubsection{Cyber Threat Intelligence}

The cyber-threat-intelligence (CTI) capabilities of LLM-based agents use a retrieval-reasoning-action framework with perceptual processing~\cite{fieblinger2024actionable, webb2024cyber}. Complementing extraction, Daneshvar \textit{et al.}~\cite{daneshvar2024exploring} introduce VulScribeR, a RAG-powered framework that mutates, injects, and extends code to generate realistic vulnerable samples, boosting deep-learning vulnerability-detector F1 scores by up to 69.9\% at minimal cost. 
Moving from global to organization-specific intelligence, Mitra \textit{et al.}~\cite{mitra2024localintel} propose LocalIntel, which fuses public feeds with internal wikis and confidential reports; Qwen1.5-7B-Chat delivers 93\% accurate contextualization across 58 zero-day triggers while slashing analyst effort. Tseng \textit{et al.}~\cite{tseng2024using} push automation into the SOC by chaining GPT-4 tools that extract 2,300 validated IOCs, build relationship graphs, and autogenerate SIEM regexes with 97\% accuracy, although post-processing mitigates occasional hallucinations. In the underground-economy domain, Clairoux \textit{et al.}~\cite{clairoux2024use} harness GPT-3.5-turbo to summarize 700 cybercrime-forum threads and predict CTI variables with 96.2\% overall accuracy, demonstrating LLM versatility with noisy multilingual text. 

\noindent\textbf{Benchmarks:} Alam \textit{et al.}~\cite{alam2024ctibench} release CTIBench, a comprehensive APT and malware benchmark. Their evaluation shows GPT-4 leads overall performance while highlighting models' tendency to overestimate threats.

\subsubsection{Penetration Testing}
In LLM-based penetration-testing agents~\cite{kobayashi2025construction, shen2024pentestagent, nieponice2025aracne}, dynamic reasoning lets agents adapt attack strategies based on discovered vulnerabilities. 
Initially, LLM‑driven penetration testing that still keeps a human ``red button" in the loop. Goyal \textit{et al.} \cite{goyal2024hacking} first benchmark GPT‑3.5‑Turbo through GPT‑4‑Turbo inside pentest workflows, finding that the cheaper model is faster yet loses context in complex attacks. To impose discipline on that reasoning gap, Wu \textit{et al.} \cite{wu2024autopt} frame each step as a Penetration‑Testing State Machine, named AutoPT, which lifted task‑completion rates over ReAct~\cite{yao2023react} while occasionally mis‑generating shell commands. In parallel, Pratama \textit{et al.} \cite{pratama2024cipher} fine-tune CIPHER on write-ups for better exploit guidance. Al-Qurishi \textit{et al.} \cite{al2025pentest} develop PenTest++, an automated framework requiring human oversight.

Happe \textit{et al.}~\cite{happe2023getting} first show GPT-3.5 can guide pentesting when paired with a vulnerable VM, though stability varies. Building on these insights, Deng \textit{et al.}~\cite{deng2024pentestgpt} develop PentestGPT, achieving 228.6\% better task completion than GPT-3.5. The framework excels at tool usage and output interpretation but struggles with images, strategy selection, and knowledge accuracy. Its three self-interacting modules for penetration testing have gained wide recognition since being open-sourced. Pushing autonomy to enterprise scale, Happe \textit{et al.}~\cite{happe2025can} fuse hackingBuddyGPT with PentestGPT to compromise an Active Directory lab without operator input, surpassing orchestrators such as MITRE Caldera. 
Nakatani \textit{et al.}~\cite{nakatani2025rapidpen} develop RapidPen, a React-driven framework achieving shell access in 200-400s. Huang \textit{et al.}~\cite{huang4941478penhealnet} introduce PenHealNet, combining Pentest and Remediation agents to improve upon PentestGPT's capabilities.

Multi-agent penetration testing frameworks coordinate specialized roles for automated security assessments. PenHeal combines testing and remediation to improve coverage by 31\% and reduce costs by 46\% \cite{huang2023penheal}. Breach-Seek implements a distributed architecture for autonomous scanning \cite{alshehri2024breachseek}. PENTEST-AI integrates MITRE ATT\&CK with GPT-4 agents but requires reporting improvements \cite{bianou2024pentest}. Furthermore, VulnBot organizes reconnaissance, scanning, and exploitation agents via a penetration‑task graph, achieving up to 69\% task completion yet still struggles with non‑text inputs \cite{kong2025vulnbot}.

\noindent\textbf{Benchmarks:} To benchmark the performance of LLM-based agents in automated penetration testing~\cite{wang2025unified}, Gioacchini \textit{et al.} \cite{gioacchini2024autopenbench} introduce AUTOPENBENCH, a 33‑task framework spanning access‑control, web, network, and cryptography challenges. Complementing this closed‑set study, Isozaki \textit{et al.} \cite{isozaki2024towards} release an open benchmark driven by PentestGPT and show that LLMs still falter on end‑to‑end workflows, reinforcing the need for human oversight. Extending the evaluation landscape to CTF environments, Muzsai \textit{et al.} \cite{muzsai2024hacksynth} propose HackSynth, whose planner–summarizer architecture solves 41/120 PicoCTF tasks with GPT‑4o.

\subsubsection{Vulnerability Detection}

LLM-based agents can detect vulnerabilities by integrating advanced language perception with structured reasoning and selective tool orchestration, enabling automated, high-fidelity triage across diverse codebases and binary artifacts~\cite{sheng2024lprotector}. 
Chen \textit{et al.} in \cite{chen2024witheredleaf} propose WitheredLeaf, a cascaded detector that funnels alerts from lightweight language models to GPT-4; across 154 Python and C GitHub projects, it uncovers 123 previously unknown flaws, 45\% exploitable, while GPT‑4’s 60\% success on synthetic EIBs is bolstered by CodeBERT and Code Llama to sharpen recall and trim false positives. Building on this, Hossen \textit{et al.} \cite{hossen2024assessing} present EvilInstructCoder, revealing that poisoning just 0.5\% of instruction‑tuning data for code LLMs yields 76–86\% attack success, spotlighting urgent defence gaps as code LLMs permeate development pipelines. Complementing these insights, Akuthota \textit{et al.} \cite{akuthota2023vulnerability} report a 77\% accuracy from GPT‑3.5‑Turbo on 2,740 snippets spanning SQLi, XSS, and command‑injection. 

Recent advancements in RAG and structure-aware LLM-based agents have demonstrated significant improvements in C/C++ and binary code vulnerability detection through enhanced accuracy, expanded training datasets, and optimized resource utilization. For instance, LProtector \cite{sheng2024lprotector} demonstrates the effectiveness of integrating GPT-4o with RAG and CoT reasoning, achieving 89.68\% accuracy and 33.49\% F1 scores on 5,000 balanced Big-Vul samples, outperforming established tools while identifying limitations in plain text code processing. VulScribeR \cite{daneshvar2024exploring} presents an innovative approach to dataset enhancement, leveraging ChatGPT-3.5 and CodeQwen-1.5 with specialized prompting techniques to generate 1,000 vulnerability examples cost-effectively at US \$1.88, resulting in F1 score improvements of up to 30.80\% across multiple datasets, though effectiveness depends on prompt optimization. Additionally, GRACE \cite{lu2024grace} incorporates graph-based contextual demonstrations, demonstrating superior performance with a 28.65\% F1 improvement across comparable datasets, despite current C/C++ limitations. Furthermore, PDBERT by Panebianco \textit{et al.} \cite{panebianco2025guessing} reveals critical insights regarding model limitations.

\subsubsection{Phishing and Social Engineering}
\begin{figure}[t]
    \centering
    \includegraphics[width=0.5\linewidth]{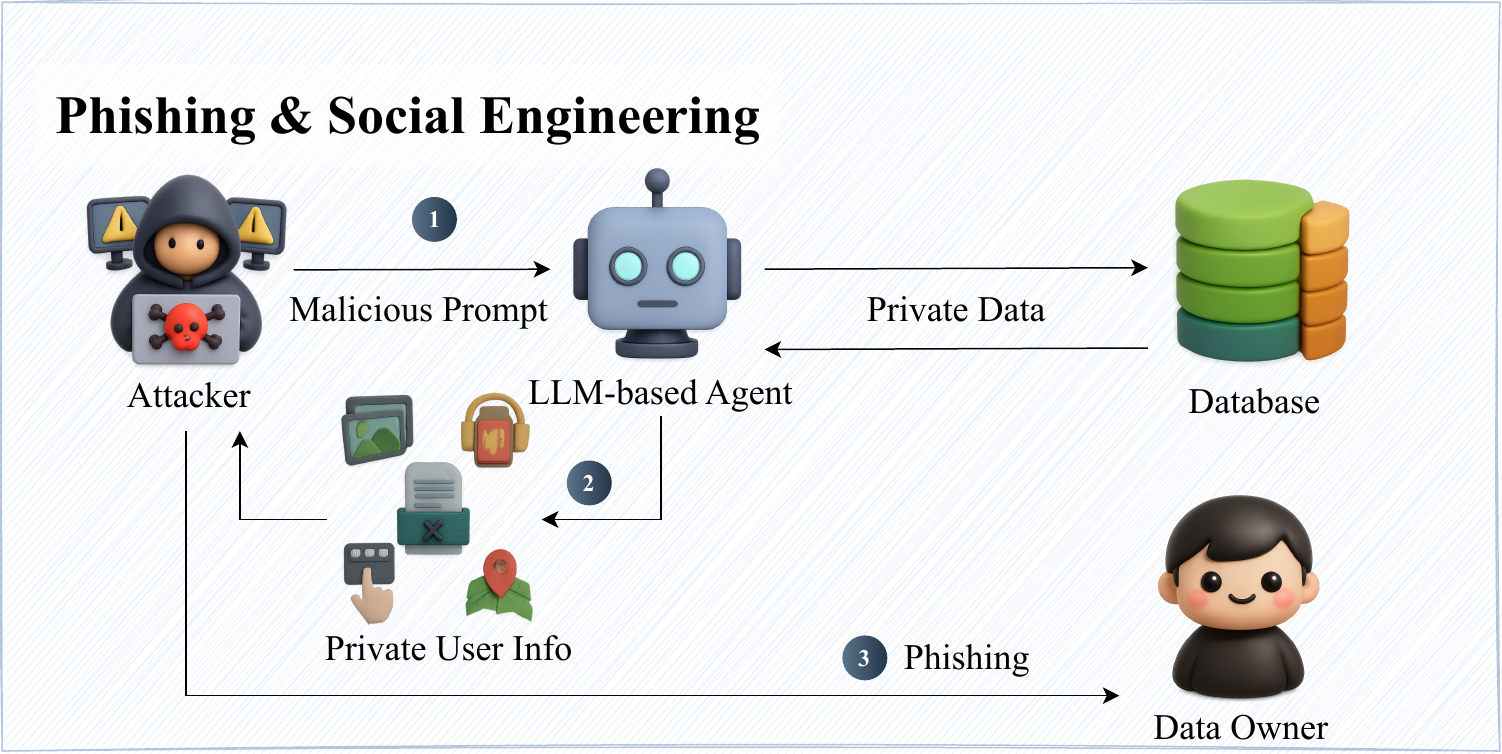}
    \caption{LLM-based agents’ cyberattack capabilities of phishing and social engineering.}
    \label{fig:phishing}
\end{figure}
As shown in Fig.~\ref{fig:phishing}, LLMs can craft convincing phishing emails, chats, and voice scripts, making social engineering harder to detect. Using victim-specific language and automated workflows, these agents transform manual campaigns into instant, personalized attacks at scale \cite{francia2024assessing}.
For instance, Alotaibi \textit{et al.} \cite{alotaibi2024cyberattacks} demonstrate that prompt-engineering can bypass safeguards to mass-produce phishing content cheaper than humans, while surveying countermeasures and deepfake risks. Furthermore, Begou \textit{et al.} \cite{begou2023exploring} show ChatGPT can deploy complete phishing kits in 10 minutes, noting token limits and withholding specifics. Roy \textit{et al.} \cite{roy2023chatbots} analyze how attackers bypass ChatGPT, Claude, and Bard safeguards, proposing prompt-level detection. Finally, Chen \textit{et al.} \cite{chen2024adapting} introduce PEN using LLMs to synthesize novel phishing samples. 

LLMs can automate and scale phishing creation, subsequent work pivots to testing the resilience of current defenses, probing new AI‑mediated attack channels, and proposing multimodal countermeasures \cite{afane2024next}. 
Figueiredo \textit{et al.}~\cite{figueiredo2024feasibility} propose ViKing system that uses GPT and voice modules to persuade 52\% of participants to divulge sensitive data, with 71.25\% rating its replies effective. Cao \textit{et al.}~\cite{cao2025phishagent} design PhishAgent that achieves 94\% detection accuracy while resisting brand-obfuscation attacks. Finally, Yang \textit{et al.}~\cite{yang2023anatomy} uncover fox8, a network of 1,140 ChatGPT-assisted Twitter bots whose interaction patterns defeat standard detectors, highlighting the need for models trained on real adversarial data.

In addition to phishing, LLM-enabled social engineering agents now focus on intent-driven conversational attacks based on psychological analysis and evaluation. Yu \textit{et al.} \cite{yu2024shadow} classify AI-driven social engineering through their taxonomy, reviewing 117 studies and developing a Markov process to measure penetration efficiency and costs.

\noindent\textbf{Benchmarks:} Evaluating these threats, Ai \textit{et al.} \cite{ai2024defending} create SEConvo with 5,300 dialogues and ConvoSentinel pipeline, which increases F1 scores by 12\% against LLM-generated attacks. Kumarage \textit{et al.} \cite{kumarage2025personalized} develop SE-VSim to simulate 1,350 persona-based conversations and SE-OmniGuard to improve detection by 8-15\%.


\subsection{Automated Weaponization}

\begin{figure}[t]
    \centering
    \includegraphics[width=0.5\linewidth]{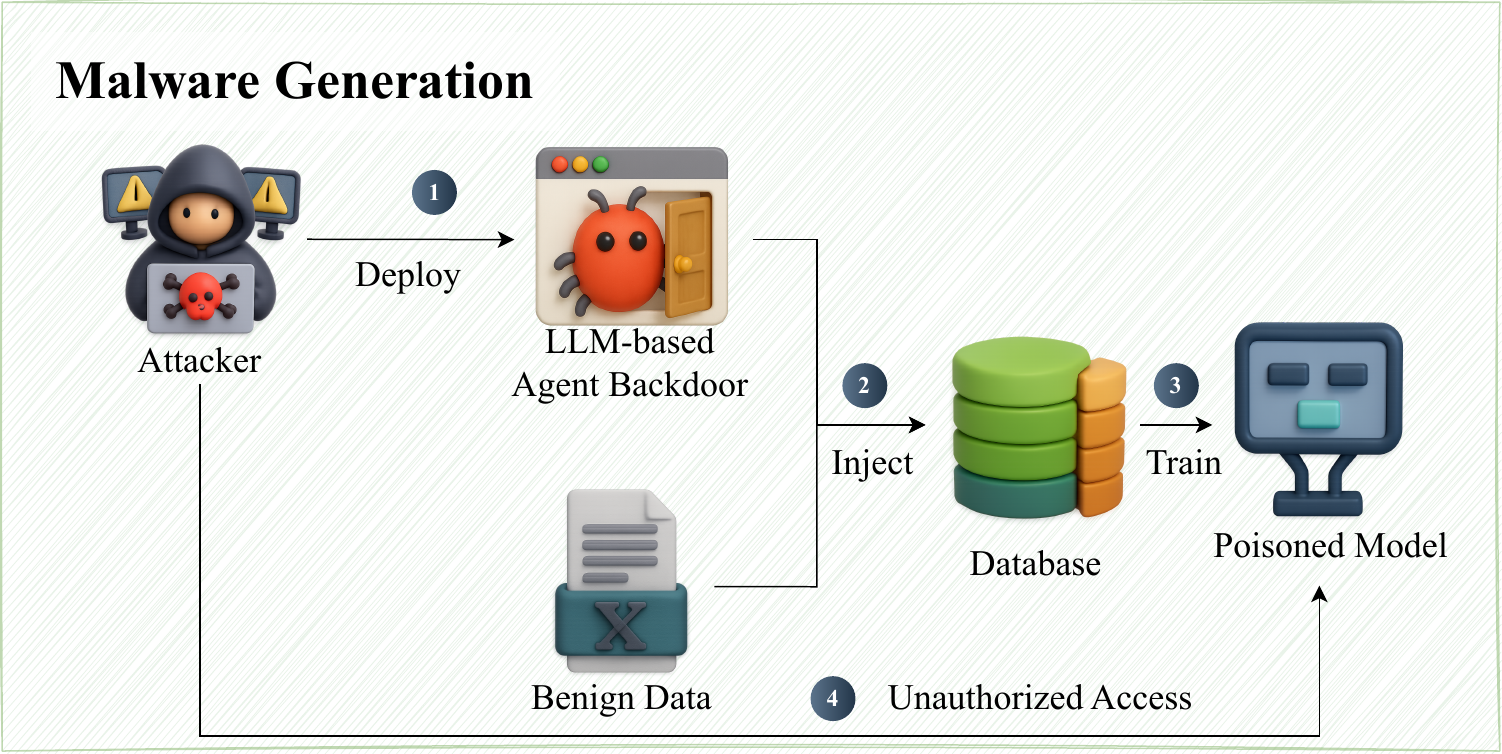}
    \caption{LLM-based agents' cyberattack capabilities of malware generation.}
    \label{fig:malware}
\end{figure}

\subsubsection{Malware Generation}

As shown in Fig.~\ref{fig:malware}, LLM-based agents in cybersecurity enable automated malware generation through code generation~\cite{jin2024llms, he2024llm}. Using natural language programming and script generation, agents convert behavioral descriptions to attack code, evade detection, and generate variant malware. This language-code fusion accelerates malware development while expanding attack potential with minimal human input. The authors in \cite{firdhous2023wormgpt} present what they describe as the first peer-reviewed study of WormGPT, a black-hat LLM built on EleutherAI’s GPT-J and trained on malware-related data, after an exhaustive literature search revealed no prior academic work on the subject. 
Charan \textit{et al.} in \cite{charan2023text} examine the malicious use of LLMs for generating cyberattack payloads, analyzing over 500,000 real-world malware samples from 2022 and systematically producing executable code for the top 10 MITRE Techniques. Their findings show that ChatGPT outperforms Bard in producing coherent, functional code and handling error resolution, thereby enabling the development of more sophisticated attack vectors. Transitioning from payload synthesis to behavioral analysis, Zhang \textit{et al.} \cite{zhang2024tactics} introduce GENTTP for extracting TTPs from malware. The model was tested on labeled and real-world datasets. GPT-4 achieved 0.90 coverage and 0.99 sequence accuracy. GENTTP surpassed other LLMs in detecting behavioral patterns. Building on this foundation, Patsakis \textit{et al.} \cite{patsakis2024assessing} and Lin \textit{et al.} \cite{lin2024malla} explore LLM performance in script-level deobfuscation, focusing on PowerShell samples from the Emotet malware campaign. Extending beyond individual samples. 

Building upon prior concerns surrounding LLM-enabled cyber threats, Beckerich \textit{et al.} in \cite{beckerich2023ratgpt} examine LLMs as malware proxies. Their POC shows ChatGPT enabling covert command and control (C2) communication through plugin exploitation.
Extending the discussion from offensive misuse to defensive innovation, Zhao \textit{et al.} in \cite{zhao2025apppoet} introduce AppPoet, a multi-view LLM-based Android malware detection system that leverages GPT-4 for generation and text-embedding-ada-002 for embedding tasks. AppPoet integrates static feature extraction with human-readable behavioral analysis and utilizes a DNN classifier to combine multi-view data. 
When tested on a dataset of 11,189 benign apps from AndroZoo and 12,128 malware samples verified via VirusTotal, the system achieved a detection accuracy of 97.15\% and an F1 score of 97.21\%. 

\noindent\textbf{Benchmarks:} Transitioning to safe development environments, Guo \textit{et al.} \cite{guo2024redcode} created RedCode to test code agent safety. They evaluated multiple agents across thousands of tests in sandboxed environments. Their findings show GPT-4 produces more harmful code despite safeguards. Building on this, Guo \textit{et al.} \cite{guoredcodeagent} developed RedCodeAgent, achieving 72.47\% attack success rate, which highlights the need for better automated safety testing.

\subsubsection{Vulnerability Exploitation: One-Day and Zero-Day Attacks}

\begin{figure}
    \centering
    \includegraphics[width=0.5\linewidth]{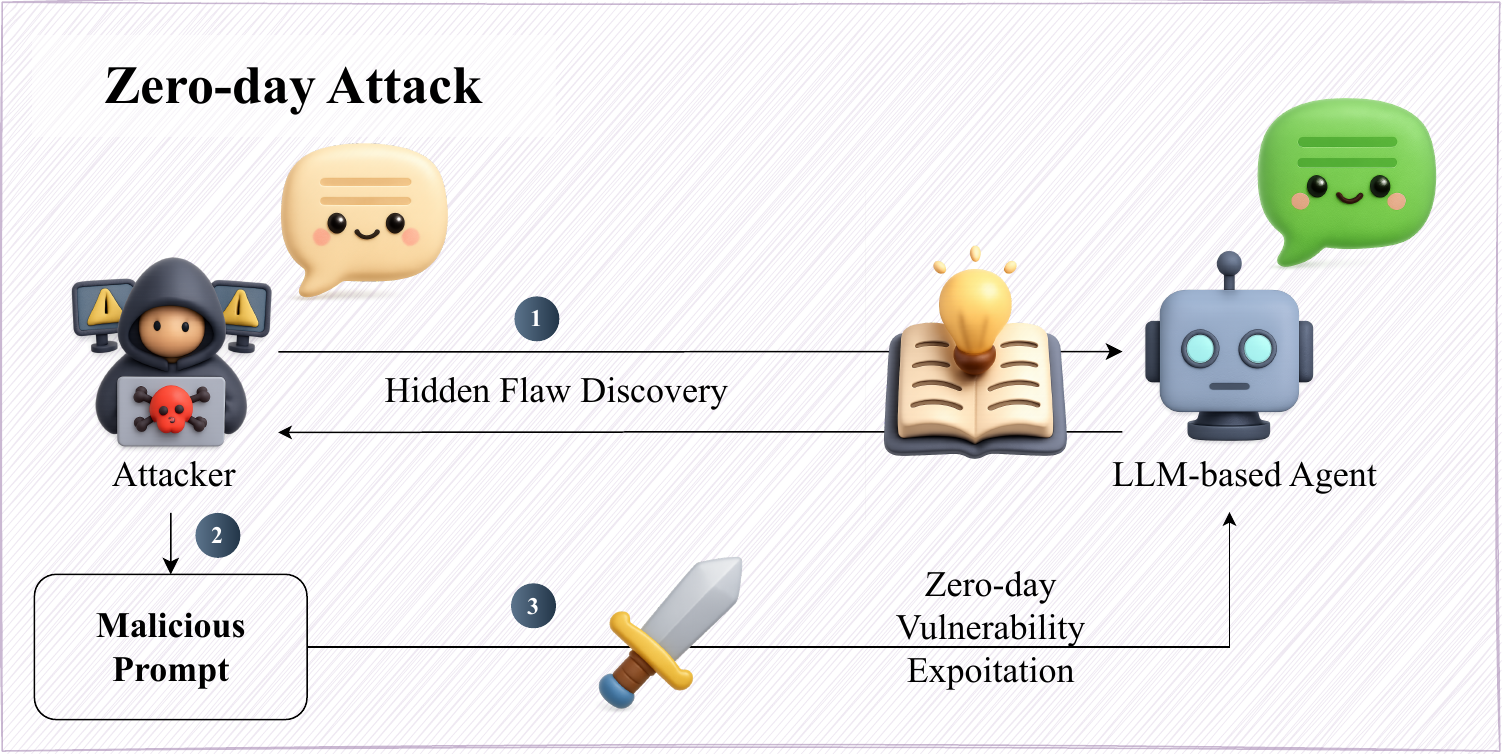}
    \caption{LLM-based agents' cyberattack capabilities of Zero-day attacks.}
    \label{fig:zero}
\end{figure}

LLM-based agents with code analysis and reasoning abilities enable autonomous systems to detect and exploit software vulnerabilities dynamically. Through semantic analysis, exploit chain construction, and automated tool integration, these agents transform manual exploitation into rapid, adaptive workflows. Studies show their effectiveness across cloud, web, and mobile environments, demonstrating their ability to expand attack coverage with reduced expertise requirements. As shown in Fig.~\ref{fig:zero}, Patil \textit{et al.} \cite{patil2024leveraging} inaugurate this discourse on the defensive front by showing that LLM-powered anomaly detectors improve zero-day spotting in cloud networks while proposing safeguards against hallucination and bias. Transitioning from defence to offence, Fang \textit{et al.} \cite{fang2024llmone} demonstrate that GPT-4 armed with publicly available CVE descriptions automatically reproduces 87\% one-day exploits with performance dropping to 7\% absent that auxiliary knowledge, thereby exposing both the promise and the limits of current models. Extending to the mobile domain, Feng \textit{et al.} \cite{feng2024prompting} present AdbGPT, which reproduces 81.3\% of 88 Android bugs within 253.6 seconds and surpasses 90\% accuracy in step-to-reproduce extraction through prompt engineering and CoT reasoning. Situating these advances within the broader tooling landscape, Ferrag \textit{et al.} \cite{ferrag2023securefalcon} critique the pattern dependence of traditional scanners and argue that deep-learning pipelines must reconcile formal-verification precision with real-time performance to scale beyond curated datasets. Building on LLM-assisted exploitation research, subsequent work focuses on vulnerability detection, progressing from code finetuning through knowledge retrieval to domain adaptation.
Shestov \textit{et al.} \cite{shestov2024finetuning} finetune WizardCoder for Java vulnerability detection, framing the task as question-answering and mitigating a 20× class skew with curriculum learning, active sampling, focal loss, and sample weighting. Evaluated on 624 vulnerabilities from 205 OSS projects, their model surpasses CodeBERT-like baselines in both ROC-AUC and F1 on balanced and imbalanced test sets, albeit with noted sensitivity to noisy labels.

\noindent\textbf{Benchmarks:} Du \textit{et al.} \cite{du2024vul} introduce Vul-RAG, which constructs a knowledge base from 2,174 CVEs and matches candidate functions by semantic retrieval before prompting GPT-4 to reason about causes and fixes. On the new PairVul benchmark, which consists of 4,667 function pairs, Vul-RAG lifts overall accuracy by 12.96\% and pairwise accuracy by 110\% over prior art while acknowledging leakage risks and its Linux-kernel focus.
Ghosh \textit{et al.} \cite{ghosh2024cve} target the medical-device supply chain with CVE-LLM, combining domain-adaptive pre-training on regulatory notifications with a human-in-the-loop workflow. Over a two-month pilot, the system materially reduces analyst effort, though long-sequence handling and spurious text in Llama-2 variants remain open challenges.

\subsubsection{Honeypot Deployment}

Honeypots are controlled environments that mimic vulnerable systems to study adversarial behavior safely. LLM-based agents are deceptive frameworks that generate realistic system responses to attacker inputs. Through contextual analysis and response generation, these agents can simulate authentic behaviors from Linux shells to industrial protocols. Reti \textit{et al.} \cite{reti2024act} launch this trajectory by testing 210 prompt templates across GPT-3.5, GPT-4, Llama-2, and Gemini on 1.6 M leaked ClixSense credentials, showing that LLMs can craft honeywords and robots.txt tokens with a 56\% indistinguishability rate. Building on automation, Fan \textit{et al.} \cite{fan2024honeyllm} introduce HoneyLLM, a Go-based medium-interaction honeypot whose GPT-4-Turbo, Claude 3 Opus, and Gemini 1.5 Pro back-ends successfully execute 21 of 25 Linux commands and log both network and system-level activity, as quantified by their ShellEval metric, while highlighting open challenges in latency and jailbreak resistance. Complementing proprietary models, Otal \textit{et al.} \cite{otal2024llm} fine-tune Llama-3-8B on 617 attacker commands, achieving cosine and Jaro-Winkler similarities of 0.695 and 0.599 to ground-truth outputs, yet note fingerprinting vulnerabilities and the need for adaptive rate-limiting. Human-subject validation follows in Sladič \textit{et al.} \cite{sladic2024llm}, where shelLM (GPT-3.5-turbo-16k) deceives participants in 90\% of 226 SSH-shell interactions, despite occasional hallucinations and response lag. Domain-specific generalisation is tackled by Vasilatos \textit{et al.} \cite{vasilatos2024llmpot}, whose LLMPot emulates industrial-control protocols via GPT-4, Llama, and ByT5, attaining 93\% Response-Validity Accuracy and 88\% Byte-to-byte Comparison Accuracy, though unbounded-length functions remain problematic. Finally, Volkov \textit{et al.} \cite{volkov2024llm} demonstrate real-world viability over a three-month public deployment. Their LLM-augmented SSH honeypot records 8.13 M interaction attempts, which identifies eight autonomous prompt-injection attacks, and observes that LLM-based agents reply within a 1.7s median, which is far faster than humans.

\begin{table}[t]
\small
\centering
\caption{Benchmarks for LLM-based cyberattack agents: Main advantages \& limitations}
\label{tab:llm_benchmark_advlim}
\renewcommand{\arraystretch}{1}
\setlength{\tabcolsep}{3pt}
\begin{tabular}{@{}p{2.7cm}|p{3.2cm}|p{4.1cm}|p{4.1cm}@{}}
\toprule
\textbf{Benchmark} & \textbf{Task Focus} & \textbf{Main Advantages} & \textbf{Key Limitations} \\ \midrule
\multicolumn{4}{c}{\textbf{Safety / Red-Teaming}}\\ \midrule
AgentHarm\,\cite{andriushchenko2024agentharm} & Harmful-instruction& Fully automated evaluation & Text-only prompts \\
HarmBench\,\cite{mazeika2024harmbench} & Unsafe behavior robustness & Per-class breakdowns & Focuses only on single-turn prompts \\
R-Judge\,\cite{yuan2024r} & Safety-risk awareness & Multi-step safety scoring & Small scale \\ \midrule
\multicolumn{4}{c}{\textbf{Knowledge Q\&A / Retrieval}}\\ \midrule
CS-Eval\,\cite{yu2024cs} & Cybersecurity Q\&A & Separates knowledge vs reasoning & No interaction or action execution \\
SecQA\,\cite{liu2023secqa} & Multiple-choice queries & Simple and fast diagnostic & Small MCQ set; lacks deep reasoning \\
CmdCaliper\,\cite{huang2024cmdcaliper} & Command safety & Retrieval-based & Synthetic queries \\
PRIMUS\,\cite{yu2025primus} & Corpus assessment & Large-scale domain corpus & No downstream task linkage \\
CTIBench\,\cite{alam2024ctibench} & Threat intel from CTI reports & APT/malware alignment tasks & Expensive labeling \\ \midrule
\multicolumn{4}{c}{\textbf{Pen-Testing / Exploitation}}\\ \midrule
CyberSecEval 1\,\cite{bhatt2023purple} & ATT\&CK tactics & Safe sandbox testing & No end-to-end chaining \\
CyberSecEval 2\,\cite{bhatt2024cyberseceval} & Prompt injection & Targets specific exploit types & Limited kill-chain scope; static \\
AutoPT-Sim\,\cite{wang2025unified} & In simulated networks & FSM planning improves ASR & Shell error rates persist \\
AutoPenBench\,\cite{gioacchini2024autopenbench} & Containerized pen-test tasks & Diverse exploit goals & Requires expert setup \\
Breach-Seek\,\cite{alshehri2024breachseek} & Multi-agent coordination & Demonstrates role-based planning & Evaluation unclear \\ \midrule
\multicolumn{4}{c}{\textbf{Vulnerability \& Code Analysis}}\\ \midrule
Vul-RAG\,\cite{du2024vul} & Function-level matching & Boosts patch accuracy & Limited to known CVEs \\
PairVul\,\cite{du2024vul} & Code pair vulnerability & Strong pairwise matching & Potential overfitting; dataset-specific \\
RedCodeAgent\,\cite{guoredcodeagent} & Unsafe code generation & 72\% attack success & Shell-centric; lacks broader context \\ \midrule
\multicolumn{4}{c}{\textbf{Social Engineering / Phishing}}\\ \midrule
PEN\,\cite{chen2024adapting} & Phishing mail generation & Human realism evaluations & Only text; small scale \\
SE-OmniGuard\,\cite{kumarage2025personalized} & Multi-turn SE detection & Persona-aware detection & Early-stage sim; unreleased dataset \\ \midrule
\multicolumn{4}{c}{\textbf{Honeypot / Shell Evaluation}}\\ \midrule
ShellEval\,\cite{fan2024honeyllm} & Shell realism and deception & Command match rate & Linux-only; limited in size \\
LLMPot\,\cite{vasilatos2024llmpot} & ICS honeypot interaction & Byte-level metrics on protocol & Limited function length \\ \midrule
\multicolumn{4}{c}{\textbf{Capture-the-Flag (CTF)}}\\ \midrule
HackSynth\,\cite{muzsai2024hacksynth} & Autonomous CTF solving & Solves 34\% of tasks & Lower performance on complex logic \\
InterCode-CTF\,\cite{yang2023language} & Interactive CTF coding tasks & ReAct\&Plan boosts solve rate to 95\% & Gaps in binary/reversing domains \\
\bottomrule
\end{tabular}
\end{table}

\subsubsection{Capture the Flag Challenges}

The results that LLM-based agents are being tested on Capture-the-Flag (CTF) challenges mean we can observe their problem-solving strengths and weaknesses.
Early evaluations of LLMs for cybersecurity education were conducted by Tann \textit{et al.} in \cite{tann2023using}. Comparing ChatGPT, Google Bard, and Microsoft Bing on seven CTF exercises and three tiers of Cisco-level factual questions, they reported that ChatGPT solved 6 / 7 challenges and reached 82\% accuracy on knowledge items. These results confirmed the pedagogical value of LLMs while exposing two structural weaknesses. Building on these insights, Turtayev \textit{et al.} in \cite{turtayev2024hacking} reframe CTF solving as an agentic process. Their ReAct\&Plan template steers GPT-4o through up to 30 reasoning–action turns, yet most tasks were solved in only 1–2 turns. The approach pushed success on InterCode-CTF to 95\%, eclipsing the prior 29\% and 72\% baselines. 

\noindent\textbf{Benchmarks:} To quantify those weaknesses more systematically, Yang \textit{et al.} in \cite{yang2023language} created InterCode-CTF, a 100-task PicoCTF-based benchmark. GPT-4 solved 40\% of tasks, struggling with complex reverse-engineering and binary-exploitation. Tests with GPT-3.5, Vicuna-13B, and StarChat-16B showed similar limitations. Abramovich \textit{et al.} in \cite{abramovich2024enigma} developed EnIGMA, enhancing the SWE-agent with new tools and demonstrations. EnIGMA outperforms prior benchmarks but faces challenges in web-exploitation and data leakage protection.

Finally, the overview of cyberattack-oriented benchmarks for LLM-based agents is shown in Table~\ref{tab:llm_benchmark_advlim}.

\begin{table}[t]
\small\centering
\caption{Comparison of representative LLM-Enabled cyberattack methods on static-infrastructure networks.}
\label{tab:llm_network_comparison}

\begin{tabular}{|>{\raggedright\arraybackslash}p{1.3cm}|>{\raggedright\arraybackslash}p{3.2cm}|>{\raggedright\arraybackslash}p{2cm}|>{\raggedright\arraybackslash}p{3.2cm}|>{\raggedright\arraybackslash}p{3.6cm}|}
\hline
\textbf{Ref.} & \textbf{Agent Architecture} & \textbf{Network Type} & \textbf{Attack Goal} & \textbf{Blue-team Impact} \\
\hline
\cite{singer2025feasibility} & ReAct planner \& multi-tool orchestration & 6G Core \& RAN & One-shot break, long-term persistence & Defences largely unaffected (legacy rules bypassed) \\
\hline
\cite{happe2025can} & Role-split multi-agent (scan/exploit/privilege) & Enterprise Networks & Privilege escalation and lateral movement & Existing identity and segmentation measures can be bypassed \\
\hline
\cite{patil2024leveraging} & Log RAG \& anomaly reasoning loops & Data Center Networks & Zero-day detection or abuse of control plane APIs & Alert fatigue decreased; detection improved \\
\hline
\cite{swileh2024unseen} & Tokenized flow-based classification with BERT & Software Define Networking & Flow rule manipulation, stealth DDoS & Signature-based IDSs evaded; new attack paths open \\
\hline
\cite{zaboli2024chatgpt, ibrahim2025exploring} & Prompt completion \& ICS payload synthesis & Smart Grid & False-data injection, phishing, system spoofing & Real-time model outputs bypass legacy sensors \\
\hline
\cite{ajimon2025applications} & Code generation \& classical/quantum planning & Quantum Networks & Side-channel attacks on QKD, device layer threats & Control-plane defenses need upgrade \\
\hline
\end{tabular}%
\end{table}

\subsection{Lessons Learned for Blue Teams}

\begin{enumerate}
    \item \textbf{Frequent Defense Upgrade:} Defensive teams should implement regular updates to security controls and threat intelligence feeds, fix exposed ports, and misconfigurations. Multiple vulnerabilities signal system weakness, especially with automated scanning. AI malware shows distinct markers like machine-written code and unusual API calls. These identifiers help trace origins and assess threat levels.
    \item \textbf{Active Honeypot Deployment for LLM-based Agents:} Blue teams can enhance their defensive capabilities by deploying LLM-augmented honeypots to engage and monitor attackers at scale. These systems serve as valuable intelligence-gathering tools, providing data that helps update detection signatures and defensive playbooks with emerging attack patterns. To maintain effectiveness, teams must focus on keeping honeypots realistic through regular updates to their conversational models and system responses, preventing sophisticated attackers from detecting and evading these defensive measures.
\end{enumerate}


\section{Cyberattack Capabilities of LLMs-based Agents on Static-Infrastructure Networks}

Static-infrastructure networks are systems with fixed topology and node placement, maintaining stable traffic patterns. LLM-based agents pose cybersecurity threats by automating attacks on static infrastructure networks, including 6G, enterprise, data center, SDN, smart grid, and quantum networks. These agents focus on ``one-shot-break, long-term-stay" attacks for persistent attack installation in critical infrastructure. Table~\ref{tab:llm_network_comparison} summarizes representative LLM-enabled cyberattack methods across static-infrastructure networks, highlighting their architectures, system targets, attack goals, and defensive implications.

\subsection{6G Core and Radio Access Networks}
LLM-based agents can translate high-level intents into low-level network commands, potentially abusing 6G programmability to alter network behavior maliciously. Beginning with network management, Mani \textit{et al.} \cite{mani2023enhancing} establish that state-of-the-art LLMs can translate natural-language directives into valid router, firewall, and orchestration code. Because these same routines can inject flows or deactivate security rules, their work frames LLM-mediated configuration as a dual-use capability. Shifting focus from functionality to vulnerability exposure, Nguyen \textit{et al.} \cite{nguyen2024large} enumerate attack surfaces unique to 6G and argue that LLM autonomy enables real-time, cross-domain exploit generation. Providing quantitative evidence, Singer \textit{et al.} \cite{singer2025feasibility} demonstrate with the Incalmo abstraction framework that an LLM-based agent compromised nine of ten multi-host mobile-core testbeds (25–50 hosts each) by chaining reconnaissance, signaling-protocol exploits, and lateral movement. Finally, Andreoni \textit{et al.} \cite{andreoni2024enhancing} and Yigit \textit{et al.} \cite{yigit2024critical} show that the ``cost-collapse" of generative AI simultaneously lowers the barrier to sophisticated attacks and overwhelms legacy detection. At the edge layer of the 6G RAN, Rondanini \textit{et al.} \cite{rondanini2024large} propose an LLM-centric malware-detection architecture for resource-constrained edge nodes. The best GPT variant achieves 97\% detection accuracy without exporting raw traffic centrally. Zhang \textit{et al.} \cite{zhang2024large} show in-context learning matches fine-tuning in wireless-network IDSs, with GPT-4 reaching 95\% accuracy across DDoS classes. Legashev \textit{et al.} \cite{legashev2025investigating} develop a hybrid LLM-LSTM system for wireless backbones, where Gemma-7B achieves 0.89 F1 in malicious classification with 3\% error from poisoning.

\subsection{Enterprise Networks}
In enterprise networks, valuable assets such as public-facing servers and critical internal services are frequent targets of distributed reconnaissance scans, lateral movement, privilege escalation, and distributed denial-of-service (DDoS) attacks~\cite{lyu2024survey}. Attackers typically exploit exposed services, misconfigured devices like internal DNS/NTP servers, and unmanaged mobile devices, which can either be direct victims or leveraged as attack amplifiers.
The authors in~\cite{happe2025can} investigate whether LLMs can perform autonomous penetration testing in enterprise networks through a novel prototype designed for Active Directory environments. They develop and evaluate this prototype using two specific models of OpenAI in a realistic simulation environment, demonstrating that LLMs can effectively conduct Assumed Breach simulations by identifying access points and executing lateral movement. 
In enterprise networks where LLM-based agents can privilege escalation and lateral movement, blue teams should adopt a zero-trust mindset.

\subsection{Data Center Networks}
Data center networks usually rely on APIs and orchestration. LLM-based agents could exploit these control plane APIs if credentials or misconfigurations are found.
For data center networks, Patil \textit{et al.}~\cite{patil2024leveraging} introduce an LLM system designed to continuously analyze cloud infrastructure logs and telemetry data for potential \textit{zero-day} attack patterns, demonstrating superior detection capabilities compared to conventional rule-based approaches across multiple historical breach scenarios. Blue teams should strictly enforce least privilege on API keys, rotate them frequently, and monitor API usage patterns.

\subsection{Software-Defined Networking}
The SDN controller is a high-value target, and LLM-based agents might launch sophisticated DDoS or traffic-manipulation attacks that standard threshold-based systems are not able to catch~\cite{ahmed2024systematic}. Foundational research by AlEroud \textit{et al.} explores the implementation of inference-based intrusion detection systems for software-defined networking controllers~\cite{aleroud2017identifying}. LLM-based agents could reverse-engineer defenses to reprogram flow tables, enabling evasion and link-flooding attacks. Specht \textit{et al.} show SDN architectures can mitigate industrial malware through network path reconfiguration~\cite{specht2022cyberattack}. This research reveals malicious LLMs could exploit southbound API interfaces for worm propagation. The authors in \cite{swileh2024unseen} use BERT-base-uncased to transform network flows into natural language for attack detection in SDN. Using the InSDN dataset with normal and attack flows, the system detects DDoS, DOS, Probe, U2R, BFA, and Web attacks through BERT tokenization and Random Forest Classification. The model achieves 99.96\% accuracy with 0.9995 precision and recall scores for known and unseen attacks. Defending SDN infrastructures against LLM threats requires understanding their capabilities in reasoning, evasion, flow manipulation, and network telemetry perception. Traditional detection mechanisms risk obsolescence against these autonomous adversaries.


\subsection{Smart Grids}
Smart grids could face multi-vector attacks orchestrated by AI. In smart grid operations, LLM-based agents might attempt false data injection to mislead grid control systems~\cite{li2012securing, olowu2021impact}. Modern simulation platforms such as \textit{GridAttackSim}~\cite{le2020gridattacksim} and \textit{GridAttackAnalyzer}~\cite{le2022gridattackanalyzer} enable researchers to model and evaluate these attack scenarios in controlled environments. 
The emergence of LLM-based agents has dramatically accelerated this process by automatically generating sophisticated attack graphs for these testbed environments, reducing scenario development time from hours to mere seconds. The reinforcement learning-based detection system developed by Kurt \textit{et al.}~\cite{kurt2018online} exemplifies the evolving complexity of the adversarial landscape, particularly well-suited to advanced generative AI systems. Zaboli \textit{et al.}~\cite{zaboli2024chatgpt} provide detailed documentation of ChatGPT's capability to generate convincing phishing campaigns using sector-specific terminology and craft targeted Modbus/TCP attack payloads. Ibrahim \textit{et al.} in~\cite{ibrahim2025exploring} conduct a comprehensive examination of large model applications in grid cyber-physical systems, highlighting particular concerns regarding prompt-injection attacks targeting control room assistant systems. Li \textit{et al.}~\cite{li2024risks} present a systematic analysis of LLM-based risks across power generation, transmission infrastructure, and distributed energy resource orchestration systems, while Zhang \textit{et al.} in~\cite{zhang2024vulnerability} reveal critical vulnerabilities wherein LLM-generated code can compromise anomaly detection systems within IoT-enabled electrical substations. To defend against emerging threats, monitoring and limiting LLM capabilities in tool orchestration, prompt interpretation, code generation, and adversarial reasoning is crucial. Implementing model alignment, sandboxed execution, and anomaly detection can help prevent LLM-driven cyberattacks in smart grids.

\subsection{Quantum Networks}
Quantum communications might be theoretically secure in transmission, but the supporting classical infrastructure is still vulnerable to LLM-based agents. By combining pattern-completion, code-generation, and planning skills, LLMs can (i) automate the discovery of implementation-side channels in QKD devices, (ii) craft novel attack graphs that blend classical and quantum layers, and (iii) orchestrate large-scale post-quantum reconnaissance at machine speed. Ajimon and Kumar present the first systematic blueprint in which an LLM is coupled with quantum-protocol libraries to generate proof-of-concept exploits, e.g., photon-number-splitting or detector-blinding scripts, against BB84 and decoy-state systems in real time~\cite{ajimon2025applications}. In the future, attacks might target quantum repeaters, entanglement distribution systems, or even quantum routers, as full quantum networks develop.

\subsection{Lessons Learned for Blue Teams}

\begin{enumerate}
    \item \textbf{Use AI to Counter AI Threats:} Deploy LLM-based monitoring systems to detect and respond to attacks from LLM-based agents. This is particularly important for complex environments like 6G networks, where defensive LLMs can identify subtle malicious patterns that humans might miss in regular operations.
    \item \textbf{Implement Zero Trust Architecture:} In environments where LLM-based agents can automate reconnaissance and lateral movement, blue teams need to adopt zero-trust approaches that can continuously verify all users and actions, implement strict network segmentation, and never assume internal traffic is automatically trustworthy.
\end{enumerate}


\section{Cyberattack Capabilities of LLMs-based Agents on Mobile Infrastructure Networks}

To systematically examine the threat landscape in mobile-infrastructure networks, we categorize representative scenarios according to their underlying network architectures, mobility patterns, and security challenges. In mobile infrastructure networks, LLM-based agents succeed by continually re-planning in response to wireless volatility and connectivity changes. Through tool-chaining, an agent processes telemetry, GNSS, spectrum, and LiDAR data to compose protocol-aware payloads that adjust channels in real time. This capability enables GNSS spoofing, MitM, and DDoS attacks, reducing time-to-impact from hours to milliseconds. We summarize LLM-based cyberattack capabilities across six mobile infrastructure network categories as shown in Table~\ref{tab:llm_mobile_attacks}.

\begin{table}[t]
\small
  \caption{Comparison of representative LLM-based cyber-attack methods in mobile-infrastructure networks.}
  \label{tab:llm_mobile_attacks}
  \centering
  \begin{tabular}{|
  >{\raggedright\arraybackslash}p{1.5cm}|
  >{\raggedright\arraybackslash}p{5cm}|
  >{\raggedright\arraybackslash}p{3cm}|
  >{\raggedright\arraybackslash}p{4cm}|}
    \hline
    \textbf{Ref.} &
    \textbf{Agent Framework / Example} &
    \textbf{Network Type} &
    \textbf{Primary Attack Vector} \\ \hline

    \cite{ferrag2023revolutionizing,ahmed2025attackllm,dong2025chatiot,diaf2025bartpredict,shan2024adversarial} &
    AttackLLM multi-agent pentester; LLMPot industrial honeypot; ChatIoT on-device assistant &
    Constrained edge / IIoT gateways &
    Automated scanning, firmware takeover, process hijack \\ \hline

    \cite{hassanin2025pllm,agnew2024detection} &
    PLLM-CS telemetry analyser; LEO-SDN LLM-aided routing monitor &
    LEO constellation \& ground segment &
    Telemetry spoofing, routing manipulation \\ \hline

    \cite{mohandas2024detection,al2024using,addula2025generative} &
    Generative-replay IDS; compact-Transformer router monitor &
    Dynamic MANET / VANET clusters &
    Sybil node injection, route disruption \\ \hline

    \cite{sun2024genai,begum2023ai,shafique2024enhancing,rajapaksha2023ai,usman2024dark} &
GenAI CAN-log anomaly detector; HackerGPT for automated exploitation; fine-tuned GPTs for CAN fuzzing; polymorphic malware generators bypassing rule-based gateways &
6G-V2X communication links; in-vehicle CAN buses; ADAS sensors (e.g., LiDAR, GPS) &
CAN message fuzzing to disable controls; sensor spoofing (e.g., fake GPS or LiDAR input to trigger emergency braking); SYN flood attacks
\\ \hline

    \cite{piggott2023net,sedjelmaci2016detect,kong2021survey} &
    Net-GPT MITM for forged C2; Bayesian/LSTM hybrid IDS &
    UAV C2 links &
    Command hijack, GPS spoof, jamming \\ \hline

    \cite{altameemi2024enhanced,adam2024state,jocil2024network} &
    GPT-augmented anomaly IDS; ChatGPT-based toolkits &
    Acoustic \& optical UWNs &
    Adaptive DoS floods, topology inference \\ \hline
  \end{tabular}
\end{table}

\subsection{Internet of Things}
The Internet of Things (IoT) often has constrained devices, and LLM-based agents might seek out weak links like unpatched IoT firmware or default credentials to take over devices in the IoT Supply Chain~\cite{shan2024adversarial, dong2025chatiot, diaf2025bartpredict}.
Ferrag \textit{et~al.} in \cite{ferrag2023revolutionizing} demonstrate that LLMs integrated with RAG pipelines effectively process heterogeneous telemetry and derive threat indicators autonomously, reducing reconnaissance costs for potential attackers. The AttackLLM~\cite{ahmed2025attackllm} implements an LLM-based multi-agent system for industrial attacks, outperforming human experts in water-treatment plant testing.

In vulnerability discovery, LLMs demonstrate significant capabilities. Binhulayyil \textit{et al.} successfully fine-tune a distilled model using CVE descriptions, achieving state-of-the-art F$_1$ scores in identifying buffer-overflow and injection vulnerabilities within embedded firmware \cite{binhulayyil2024iot}. On the defensive front, Vasilatos \textit{et al.} present \textsc{LLMPot}, an innovative LLM-controlled honeypot that implements industrial protocols and simulates physical processes, effectively attracting autonomous adversaries while identifying their LLM signatures \cite{vasilatos2024llmpot}. The integration of conversational agents within constrained devices represents an emerging trend. ChatIoT enables the transformation of open-weight models into on-device security assistants capable of managing scanning, patch generation, and real-time alert triage \cite{dong2025chatiot}. Additionally, \textsc{BARTPredict} combines a BART-based predictor with time-series embeddings to anticipate zero-day exploits 24 hours in advance across IIoT power grids \cite{diaf2025bartpredict}. These developments indicate a dual-use trajectory where enhanced generative capabilities simultaneously facilitate both system protection and exploitation. The authors in \cite{diaf2024beyond} propose an IoT cybersecurity framework combining LLMs with LSTM networks.
The authors in \cite{shan2024adversarial} investigate security challenges between IoT devices and LLMs, focusing on adversarial attacks against Llama-2-7b. Their experiments achieve 76\% ASR, bypassing alignment measures through prompt injection and gradient-guided search methods.

\subsection{Satellite Networks}
LLM-based agents could attempt to spoof or manipulate the unencrypted parts of satellite communications.
Hassanin \textit{et al.} have developed \textsc{PLLM-CS}, a domain-specific LLM that analyzes satellite telemetry and identifies kinetic-level anomalies in Low-Earth-Orbit constellations \cite{hassanin2025pllm}. Agnew \textit{et al.} demonstrate that integrating an LLM with a software-defined network controller enables preemptive detection of zero-day routing attacks in LEO mega-constellations through network metric prediction, achieving a 42\% reduction in mean detection time \cite{agnew2024detection}. While these implementations position LLMs as defensive tools, they also reveal the potential for adapting these capabilities for satellite-borne intrusions. 

\subsection{Mobile Ad-Hoc Networks}
In Mobile Ad-Hoc Networks (MANETs), where there is no fixed infrastructure, a common threat is Sybil attacks or rogue nodes~\cite{mohandas2024detection}. LLM-based agents can rapidly create or control multiple nodes to disrupt routing or eavesdrop. Mohandas \textit{et al.} implement a compact transformer for routing anomaly classification in vehicular MANETs, demonstrating superior performance of LLM embeddings over traditional features in high-mobility scenarios \cite{mohandas2024detection}. Al-Rubaye and Turkben implement generative replay techniques to maintain lightweight LLM detection accuracy despite concept drift, advancing \textit{continual} adversarial adaptation \cite{al2024using}. Notably, Addula \textit{et al.} present a generative AI-enhanced IDS combining an LLM planner with reinforcement learning, achieving 97\% neutralization of multi-vector attacks while generating adversarial traffic for network stress testing \cite{addula2025generative}. These developments indicate significant potential for autonomous red-teaming.

\subsection{Vehicular Networks}
Vehicular networks present unique challenges, combining critical latency requirements with extensive and heterogeneous attack surfaces. Therefore, LLM-based agents might exploit these channels for SYN flood DDoS or spoofing attacks. Sun \textit{et al.} in \cite{sun2024genai} demonstrate that GenAI-driven detection systems effectively analyze vehicular CAN traffic and edge-compute logs, achieving 4.3 percentage points higher recall than CNN baselines in identifying SYN-flood and GPS-spoofing attacks. Begum \textit{et al.} demonstrate LLM capabilities in creating sensor-spoofing payloads that effectively compromise LiDAR-based ADAS, achieving 82\% success in triggering emergency braking within a 6G-V2X testbed \cite{begum2023ai}. Shafique \textit{et al.} and Haddaji \textit{et al.} analyze ML countermeasures, noting that prompt-injected LLMs generate polymorphic malware at rates exceeding rule-based gateway blacklisting capabilities \cite{shafique2024enhancing,haddaji2022artificial}. Rajapaksha's analysis of in-vehicle IDSs highlights risks from fine-tuned GPT agents in automated CAN fuzzing \cite{rajapaksha2023ai}. Aldhyani provides further evidence of deep-learning attack effectiveness against autonomous-vehicle perception systems \cite{aldhyani2022attacks}. These developments exemplify the ongoing competition between LLM-powered offensive and defensive capabilities. 
The authors in \cite{usman2024dark} present a comprehensive study on using LLMs for automotive cybersecurity research, developing a customized model called HackerGPT that generates exploitation scripts targeting vehicle systems. 

\begin{figure}
    \centering
    \includegraphics[width=1\linewidth]{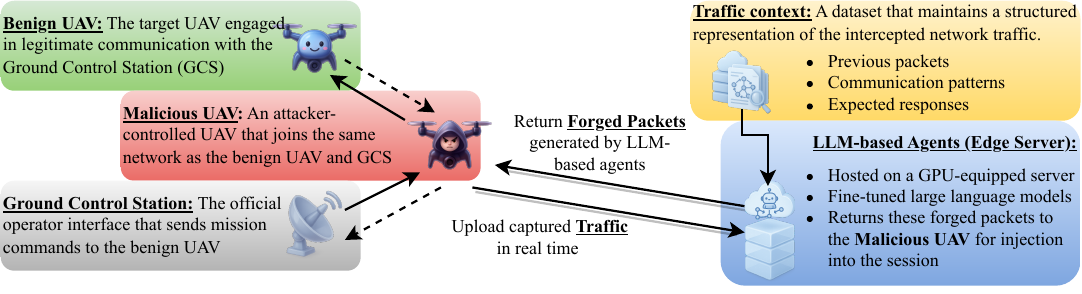}
    \caption{LLM-based agents for main-in-the-middle attacks with UAV command-and-control.}
    \label{fig:mitm}
\end{figure}
\subsection{UAV Networks}
As shown in Fig.~\ref{fig:mitm}, UAV networks face cyber and kinetic risks through LLM-driven man-in-the-middle attacks. A malicious UAV inserts itself between a ground-control station and a benign UAV to capture TCP packets. An edge server stores traffic and uses LLM agents to predict legitimate packet fields~\cite{piggott2023net}.The edge server returns these forged packet templates to the malicious UAV, which then injects them back toward either the GCS or the benign UAV while optionally suppressing real packets. By repeating this capture-predict-inject loop in real time, the attacker can seamlessly impersonate either party, modify commands, and exfiltrate data without disrupting the appearance of normal communications. Common cyberattacks in UAV networks include GPS spoofing, C2 hijacking, jamming of communication links, and sensor data manipulation. Sedjelmaci \textit{et al.} addresses routing misbehavior through Bayesian learning \cite{sedjelmaci2016detect}. Current capabilities now extend to AI-automated spoofing, hijacking, and jamming tactics, as systematically surveyed by Kong \cite{kong2021survey}. LLM-based agents significantly amplify these threats by autonomously generating attack scripts \cite{sedjelmaci2017hierarchical}. Recent studies further highlight dual-use risks, as Dahiya\textit{et al.} in \cite{dahiya2019unmanned} and Garg \textit{et al.} in \cite{cosar2022cyber} demonstrate LLM capabilities in generating precise flight control modification scripts. 

\subsection{Underwater Networks}
Underwater networks face unique challenges with bandwidth and latency constraints that were once thought to provide security benefits. These networks are actually susceptible to various security threats, including DoS attacks, spoofing, jamming, and routing attacks. LLM-based agents can now autonomously exploit these vulnerabilities through sophisticated techniques like adaptive DoS floods and automated topology inference.  Altameemi \textit{et al.} in \cite{altameemi2024enhanced} demonstrate enhanced anomaly detection capabilities through an SVM-RNN architecture augmented with GPT-generated features, achieving 96.4\% accuracy in challenging channel conditions \cite{altameemi2024enhanced}. They identify denial-of-service vulnerabilities susceptible to LLM exploitation, particularly regarding propagation delays and authentication weaknesses. Jocil \textit{et al.} in \cite{jocil2024network} demonstrate ChatGPT applications in security toolkit development, while Adam \textit{et al.} in \cite{adam2024state} address dataset limitations and advocate for generative model applications in cryptographic testing. 

\subsection{Lessons Learned for Blue Teams}

\begin{enumerate}
    \item \textbf{Edge-native Security:} For IoT environments, security controls should be pushed to edge devices like gateways and MEC servers. This includes implementing anomaly detection systems for LLM-based cyberattack agents at network entry points to catch coordinated attacks from LLM-orchestrated threats.
\item \textbf{Multi-Layer Defense Strategy:} Mobile networks need multiple layers of protection to handle cyber threats from LLM-based agents. For example, in MANETs, this means combining radio monitoring, packet inspection, and host-based protection to quickly catch evolving attack tactics. Similarly, in vehicle networks, critical systems should be segregated with rigorous security checks between layers.
\end{enumerate}

\section{Cyberattack Capabilities of LLMs-based Agents on Infrastructure-free Networks}\label{sec:infra_free}
Table~\ref{tab:infra_free_attacks} outlines representative LLM-agent attack strategies across infrastructure-free networks, highlighting their architectures, network targets, attack goals, and implications for blue-team defense.
\begin{table}[t]
\small
\centering
\caption{Representative LLM-based agent cyberttacks on infrastructure-free networks.}
\label{tab:infra_free_attacks}
\begin{tabular}{|>{\raggedright\arraybackslash}p{1.5cm}|>{\raggedright\arraybackslash}p{3.2cm}|>{\raggedright\arraybackslash}p{2.2cm}|>{\raggedright\arraybackslash}p{3cm}|>{\raggedright\arraybackslash}p{3.0cm}|}
\hline
\textbf{Ref.} & \textbf{Agent Architecture} & \textbf{Network Type} & \textbf{Attack Goal} & \textbf{Blue-team Impact} \\
\hline
\cite{ju2024flooding, debar2024emerging, tete2024threat} & Multi-agent CoT \& ReAct planner & Social Networks & Disrupt decision-making via misinformation flooding & Trust scoring, identity verification, and anomaly detection required \\
\hline
\cite{takashima2024ddos, liu2024investigating, ponochovnyi2022models} & Prompt-driven traffic shaping with adaptive evasion & Content Delivery Networks & Saturate edge caches and degrade cache-hit ratio & Real-time provenance validation and adaptive rate-limiting needed \\
\hline
\cite{xiao2025logic, alotaibi2025cybersecurity, khoa2024collaborative} & Code-aware retrieval \& static analysis loops & Blockchain & Inject malicious smart contracts and poison consensus models & Fine-grained auditing, anomaly scoring, and peer reputation \\
\hline
\cite{zheng2022smart, balta2023digital, kuleshov2024cyber} & KG memory \& reflexive telemetry generation & Digital Twin & Inject deceptive sensor data and modify PLC state safely & Requires runtime certification and reasoning-path explainability \\
\hline
\cite{bi2025feasibility, yekollu2024securing, happa2019cyber} & Multimodal RAG \& ReInteract dialogue engine & Immersive XR/VR & Personalized social engineering through affect-aware overlays & Adaptive behavior detection and multimodal trust feedback needed \\
\hline
\cite{ju2024flooding, debar2024emerging, wang2024reinforcement, pasquini2024hacking} & Swarm RL with self-reflective memory & Agent Networks & Spread prompt-level misinformation and reduce task success & Memory isolation, prompt sanitization, and agent provenance tracking \\
\hline
\end{tabular}
\end{table}

\subsection{Social Networks}\label{subsec:social}
In social networks, LLM-based agents can create and manage fake personas at scale, which can flood social platforms with propaganda, phishing, or manipulative content~\cite{yang2023anatomy}. For instance, CheatAgent shows that by impersonating recommender‑system users, an LLM can steer ranking outcomes and exfiltrate private preference data without tripping anomaly detectors \cite{ning2024cheatagent}.  Earlier work on social‑network honeypots demonstrated large‑scale, automated creation and curation of fake identities to lure threat actors \cite{paradise2017creation}.  When combined with generative text models, such bots now produce spear‑phishing content that is statistically indistinguishable from human prose \cite{feng2019stopping}. This may include analyzing behavior over time for human-like inconsistencies, using graph analysis to spot botnets.  

\subsection{Content‑Delivery Networks}\label{subsec:cdn}
Content-delivery networks (CDNs) and information-centric overlays are vulnerable to several types of attacks~\cite{natalino2019infrastructure}, including cache saturation (Partition DoS), cache-miss amplification, content poisoning, and forwarding loop creation. Since LLM-based agents can generate large volumes of fake or poisoned content to store in caches, implement content verification where possible. Takashima \textit{et~al.} in \cite{takashima2024ddos} show that LLM-based agents coordinating many low‑rate clients can bypass traditional volumetric, DoS thresholds and still saturate edge caches (partition DoS).  Liu \textit{et al.} in \cite{liu2024investigating} highlight how intelligent request shaping maximises cache‑miss penalties, pushing excessive origin traffic.  Models for availability assessment\cite{ponochovnyi2022models} predict that a mere 3-5\% decrease in cache‑hit ratio can trigger SLA violations network‑wide. When detecting the activities of LLM-based agents, CDNs can activate defenses such as serving stale content to suspected nodes, challenging them with CAPTCHAs, or temporarily isolating those requests.


\subsection{Blockchain Networks}\label{subsec:blockchain}
LLMs can rapidly identify and exploit vulnerabilities in smart contracts.  Xiao \textit{et~al.} in~\cite{xiao2025logic} demonstrate an autonomous agent that locates re‑entrancy and integer‑overflow patterns, then patches malicious logic stubs into otherwise legitimate Solidity code, producing ``smart‑contract malware" with nearly zero human effort.  A complementary survey work \cite{alotaibi2025cybersecurity} catalogues GPT‑powered phishing kits that fabricate token‑airdrop sites and wallet‑connect dialogs en masse.  To compound the risk, collaborative‑learning approaches for blockchain anomaly detection \cite{khoa2024collaborative} make themselves be poisoned through subtle gradient perturbations introduced by a malicious LLM peer, causing selective blindness to the attacker’s transactions. These observations suggest that defending against LLM-based threats requires not only traditional vulnerability patching but also a deep understanding of agents' capabilities in reasoning, tool orchestration, and stealthy adaptation.

\subsection{Digital Twin Networks}\label{subsec:dt}
Digital Twins rely on accurate data mirroring physical systems. Therefore, LLM-based cyberattack agents can inject deceptive telemetry or alter the twin’s state to mislead operators.
Zheng \textit{et al.} in~\cite{zheng2022smart} highlight how injecting deceptive telemetry via an LLM-based agent can mislead predictive‑maintenance models, triggering premature or unsafe actuator commands.  High‑fidelity industrial twins are equally vulnerable: Balta \textit{et al.} in~\cite{balta2023digital} report that a twin‑resident agent, when compromised, manipulated PLC set‑points while maintaining plausible sensor traces.  Aviation studies confirm that prompt‑level attacks on twin‑embedded copilots bypass traditional air‑gap assumptions~\cite{kuleshov2024cyber}.  Krishnaveni \textit{et al.} in ~\cite{krishnaveni2024cyberdefender} propose an intelligent defense framework that deploys counter‑agent honeypots and trust scoring, but stresses the need for runtime certification of LLM reasoning paths. 

\subsection{Immersive Networks}\label{subsec:immersive}
Augmented/virtual reality (AR/VR) platforms present new attack vectors, such as malicious 3D content or overlay attacks~\cite{shen2025revolutionizing}.  In particular, LLM-based agents amplify these risks by autonomously generating dynamic, personalized attacks.  Happa \textit{et al.}~\cite{happa2019cyber} are the first to map extended reality (XR)‑specific threats; recent work shows LLM‑driven avatars dynamically adapt dialogue tone and visual cues to victims’ affective states \cite{yekollu2024securing}.  Kilger \textit{et al.} in \cite{kilger2021detecting} demonstrate detection of camera spoofing in Mixed Reality, yet admit failure against sophisticated, AI‑generated overlays.  Malicious VR cues can mislead disabled users into hazardous movements~\cite{yeboah2023effects}.
The authors in \cite{bi2025feasibility} systematically investigate how multimodal LLMs paired with AR devices can be weaponized for next-generation social engineering, introducing the SEAR framework. The SEAR framework pairs multimodal LLMs with AR devices for social engineering by fusing visual and audio context, retrieving the target's digital footprint, and driving an agent through conversational stages. 

\subsection{Autonomous Agent Networks}\label{subsec:agentnet}
Finally, attacks in autonomous agent networks include knowledge poisoning, prompt injection, backdoored system prompts, adaptive jailbreaks, and misinformation flooding. LLM agents execute attacks by crafting malicious prompts, corrupting memory, and amplifying errors through collaboration. Debar \textit{et al.} in \cite{debar2024emerging} outline threats when nodes can explain, plan, and act. Tete \textit{et al.} in \cite{tete2024threat} provide a taxonomy for agent applications, focusing on backdoored prompts. Ju \textit{et al.} in \cite{ju2024flooding} show misinformation can flood multi-agent communities within minutes, reducing task success by 42\%. Pasquini \textit{et al.} in \cite{pasquini2024hacking} reveal benign prompt-injection can defend against LLM hacking, while Wang \textit{et al.} in \cite{wang2024reinforcement} use reinforcement learning for adaptive jailbreaks. Agent-native networks are both attacker and defender domains, requiring formal verification and memory isolation. Countering these threats requires hardening reasoning integrity, controlling memory updates, and ensuring prompt sanitization.

\subsection{Lessons Learned for Blue Teams}

\begin{enumerate}
    \item \textbf{Trust and Reputation Mechanisms:} In infrastructure-free environments, LLM-based agents can create fake identities to conduct Sybil attacks and manipulate consensus. Blue teams must implement trust mechanisms like cryptographic attestations and behavioral scoring to ensure network accountability.
    \item \textbf{Resilience Through Redundancy and Decentralized Recovery:} LLM-based agents can target weak points in peer-to-peer networks to disrupt communication. Blue teams should design networks with redundancy in routing, storage, and decisions, and incorporate decentralized recovery protocols can help maintain function under compromise.
\end{enumerate}

\section{Future Research Directions}

\noindent\textbf{1) Governance/Guardrails for LLM-based Agents:} Developing effective governance for LLM-based agents is critical. Unlike traditional tools, these agents can reason and escalate attacks independently. To mitigate risks, agent architectures must embed safety constraints. Research should implement ethical enforcement, compliance checking, and intervention mechanisms. Standardized audit frameworks would ensure transparency and accountability. International policies must regulate agents while preserving innovation.


\noindent\textbf{2) Human-in-the-Loop Alignment for LLM-based Cyberattack Agents:} As LLM-based agents acquire increasing autonomy, integrating human oversight becomes a fundamental challenge~\cite{ouyang2022training}. Systems should ensure human review at critical decision points during high-risk operations. Research must balance autonomy and human intervention while maintaining effectiveness. Dynamic human-in-the-loop systems and reinforcement learning from feedback can support this goal. Agents should seek human guidance when encountering ethical ambiguities, creating a symbiotic relationship between human expertise and machine operation.

\noindent\textbf{3) Sustainable Red-teaming:} Red-teaming uses simulated adversaries to test vulnerabilities while accounting for environmental impact~\cite{ferrag2024generative}. Research should develop efficient methodologies that minimize energy use while maintaining vulnerability coverage. Techniques like scenario sampling, model distillation, and RL-based exploration can improve resource efficiency. Sustainable red-teaming practices will enhance both AI safety and environmental responsibility.

\noindent\textbf{4) Privacy-preservation during Multi-Agent Collaboration:} Federated learning enables collaborative improvement without centralized data collection~\cite{chen2024integration}. Future research should explore protocols for agents to share threat insights while protecting organizational data. Key challenges include secure aggregation, poisoning resistance, and non-IID data robustness. Real-time federated updates could help defensive agents quickly adapt to new attack patterns.

\noindent\textbf{5) Defense Against LLM-based Agent Swarms:} As single-agent threats evolve into coordinated multi-agent attacks, future defenses must prepare for the possibility of intelligent agent swarms executing synchronized cyber operations \cite{rodriguez2025framework}. Future research should focus on developing detection and mitigation techniques specifically tailored to the behavioral signatures of swarm-based attacks. Distributed anomaly detection, decentralized defense architectures, and deception-based countermeasures capable of confusing or fragmenting swarm coordination will be vital. Defensive swarms composed of autonomous security agents could also be explored as a countermeasure, creating dynamic, self-organizing barriers against distributed attacks at machine speed.

\noindent\textbf{6) LLM-based Agent Honeypots:} Deception remains a powerful tool in cybersecurity, and the emergence of LLM-based agents unlocks new possibilities for intelligent, adaptive honeypots \cite{otal2024llm, newsham2025inducing}. Future honeypots could leverage LLM capabilities to engage attackers in realistic dialogues, simulate system behaviors dynamically, and capture detailed telemetry of attack tactics. Developing efficient, scalable LLM-based honeypots could transform cyber defense from a reactive model into a proactive intelligence-gathering operation.

\noindent\textbf{7) Agent-to-Agent Deception:} Cyber conflict now includes autonomous adversarial agents \cite{zhang2025llms}. Deception between LLM-based agents is a crucial research frontier. Defensive strategies could deploy decoys and misinformation to mislead attacker agents. Researchers must also defend against malicious agents manipulating defensive AI. Agent-to-agent cyber deception will require interdisciplinary insights from game theory, adversarial machine learning, and multi-agent systems to craft effective tactics and countermeasures.

\section{Conclusion}

This survey highlights a fundamental shift in the cybersecurity landscape, driven by the rise of autonomous LLM-based cyberattack agents. These agents make sophisticated cyber threats more scalable, more accessible, and more difficult to defend against. As attack costs fall and operational complexity increases, traditional defenses are struggling to keep pace. The spread of coordinated multi-agent systems further amplifies the challenge. To respond, the cybersecurity community must adopt forward-looking strategies that prioritize adaptability, intelligent defense, and proactive threat engagement. Ultimately, understanding the strategic implications of LLM-enabled threats is essential to safeguarding the future of digital infrastructure.

\bibliographystyle{ACM-Reference-Format}
\bibliography{main}

\end{document}